\documentclass{aa}

\usepackage{amsmath}
\usepackage{graphicx}
\usepackage[varg]{txfonts}
\usepackage{natbib}

\begin{document}

\title{Extended view on the dust shells around two carbon stars}

\author{M.~Me\v{c}ina\inst{\ref{inst1},\ref{inst5}} \and B. Aringer\inst{\ref{inst2}} \and W. Nowotny\inst{\ref{inst1}} \and M.A.T. Groenewegen\inst{\ref{inst3}} \and F. Kerschbaum\inst{\ref{inst1}} \and M. Brunner\inst{\ref{inst1}} \and H.-P. Gail\inst{\ref{inst4}}}

\institute{Department of Astrophysics, University of Vienna, T\"{u}rkenschanzstra{\ss}e 17, 1180 Vienna, Austria\\\email{marko.mecina@univie.ac.at}\label{inst1} \and Dipartimento di Fisica e Astronomia Galileo Galilei, Universit\`a di Padova, Vicolo dell'Osservatorio 3, I-35122 Padova, Italy\label{inst2} \and Koninklijke Sterrenwacht van Belgi\"{e}, Ringlaan 3, 1180 Brussels, Belgium\label{inst3} \and Institut f\"{u}r Theoretische Astrophysik, Zentrum f\"{u}r Astronomie, Universit\"{a}t Heidelberg, Albert-Ueberle Str. 2, 69120 Heidelberg, Germany\label{inst4} \and Space Research Institute, Austrian Academy of Sciences, Schmiedlstra{\ss}e 6, 8042 Graz, Austria\label{inst5}}

\date{}

\abstract
{Stars on the asymptotic giant branch (AGB) lose considerable amounts of matter through their dust-driven stellar winds. A number of such sources have been imaged by Herschel/PACS, revealing a diverse sample of different morphological types. Among them are a few examples which show geometrically thin, spherically symmetric shells which can be used to probe the mass loss history of their host stars.}
{We aim to determine the physical properties of the dust envelope around the two carbon stars U~Hya and W~Ori. With the much-improved spatial constraints from the new far-infrared maps, our primary goal is to measure the dust masses contained in the shells and see how they fit the proposed scenarios of shell formation.}
{We calculated the radiative transfer of the circumstellar dust envelope using the 1D code \textit{More of DUSTY} (MOD). Adopting a parametrised density profile, we obtained a best-fit model in terms of the photometric and spectroscopic data, as well as a radial intensity profile based on Herschel/PACS data. For the case of U~Hya, we also computed a grid of circumstellar envelopes by means of a stationary wind code and compare the results of the two modelling approaches.}
{The Herschel/PACS maps show U~Hya surrounded by a detached shell of 114\arcsec (0.12\,pc) in radius, confirming the observations from previous space missions. The dust masses calculated for the shell by the two approaches are consistent with respect to the adopted dust grain properties. In addition, around W~Ori, we detect for the first time a weak spherically symmetric structure with a radius of 92\arcsec (0.17\,pc) and a dust mass of $(3.5\pm0.3)\times10^{-6}\,\mathrm{M_\odot}$.}
{}

 
 \maketitle

\section{Introduction}
In the Milky Way, low- to intermediate-mass stars (0.8-8\,M$_\odot$) represent a significant fraction of the stellar population \citep[e.g.][]{Chabrier2003}. While the impact on the immediate surroundings may be comparatively small on the individual level, their sheer number motivates studies of their role in the cosmic cycle of matter.

When these stars ascend the asymptotic giant branch (AGB), producing energy via  hydrogen and helium shell burning, they develop a pronounced convective envelope. Following thermal pulses of the He-burning shell, which appear during the later phases of the AGB evolution \citep[TP-AGB,][]{Herwig2005}, the products of nucleosynthesis ($^4$He, $^{12}$C, $^{22}$Ne, $^{25}$Mg, and heavy isotopes produced by the s-process) are dredged up to the stellar atmosphere \citep{Karakas2014}. Consequently, the surface chemistry and the composition of the stellar envelope may change from an initially oxygen-rich to a carbon-rich mixture. 

For solar metallicity, this transition from C/O\,<\,1 (M type stars) to C/O\,>\,1 (carbon stars or C-type stars) is limited to objects with initial masses between $\approx$\,1.5 and 4$\,$M$_{\odot}$ \citep{Hoefner2018}. Stars with masses below that range do not dredge up processed elements efficiently enough for the abundance ratio to reach C/O\,>\,1. For higher mass objects, hot bottom burning destroys the produced carbon via the CNO cycle, which also leads to an  overabundance of nitrogen. In an environment, however, where enough carbon (C/O\,>\,1) reaches the upper layers of the atmosphere, carbon-bearing species, such as C$_{2}$, C$_{2}$H$_{2}$, C$_3$, CN, HCN, etc., dominate the gas phase (besides H$_2$ and CO) and partially condense into solid particles, that is -- dust. This dust primarily consists of amorphous carbon grains (amC), however, depending on the exact conditions in the formation zone, the particles could, in principle, also exhibit a different type of structure. Another quite commonly found dust species after pure carbon is SiC, a highly refractory material, detectable by its pronounced spectral feature at $\mathrm{11.2\ \mu m}$. It may serve as a seed particle, upon which other species, such as amC, can condense to effectively form larger grains. Observations indicate that MgS is also present in solid form around AGB stars, identified by a broad, loosely defined feature around $\mathrm{30\ \mu m}$ \citep{Lombaert2012}. In principle, the fundamental processes of dust formation are known, but the critical steps in the transition from the microscopic to the macroscopic regime are not yet entirely understood, and the search for the prevailing nucleation paths makes up the focus of theoretical studies \citep{Cherchneff2006,Cherchneff2012,Gobrecht2016,Gobrecht2017}.

In cooler AGB stars of type M and C, the combination of pulsation and radiation pressure on dust grains is the main driver of mass loss, through which a major fraction of the initial stellar mass is expelled into the interstellar medium (ISM), including the gas component, which typically constitutes more than 99\% of the total outflow mass. Identifying the dust species involved in this process is essential, as they show very different behaviour in their interaction with the stellar radiation field. In the case of C-rich atmospheres and, more recently, the rather intricate case of M stars, models of dust-driven winds and pulsation have been able to reproduce observed mass-loss rates and expansion velocities \citep{Eriksson2014,Bladh2019,Bladh2019a}. As a consequence of their large share of the stellar population, AGB stars are believed to greatly contribute to the dust budget in the Local Universe \citep{Zhukovska2008,Schneider2014,Nanni2018,Nanni2019}. However, as another potential contributor, the role of supernovae (SNe) in the dust production cycle is still debated \citep{Matsuura2015}. It is not clear how much of the apparently formed dust endures; moreover, grain growth in the ISM is argued to be of importance for the dust budget as well \citep[e.g.][]{DeVis2017a,DeVis2019,Zhukovska2014}.

The mass-loss rate of AGB stars is not constant and, on average, it increases as the star evolves until the AGB phase is terminated. Typical values for the stellar wind range from $ 10^{-8}$ up to $10^{-4}\,\mathrm{M_{\odot}/yr} $ \citep{Abia2001}. This long-term change may be modulated by reoccurring, relatively short ($\approx$\,100 years) periods of significantly increased mass loss, which are believed to be connected to thermal pulses \citep{Steffen2000}. These variations are imprinted in the circumstellar envelope and can be detected over the whole spectral range, depending on the distance from the stellar surface and, hence, on the temperature \citep[see, e.g.][]{Olofsson1990, GonzalezDelgado2001, Maercker2010, Kerschbaum2010}. 

In the simplest case, we can expect a radially variable, spherically symmetric density distribution. Indeed, several geometrically thin shells (also referred to as `detached shells') have been observed. For example, \citet{Olofsson2000} detected such a structure around \object{TT~Cyg} in CO line emission, clearly indicating a (rather recent) temporal increase in the mass-loss rate. Additional targets showed similar matter distributions in their environment when they were studied in scattered optical light \citep[e.g.][]{GonzalezDelgado2003,Olofsson2010,Maercker2014}. The sample was further extended and complemented by Herschel/PACS observations of far-infrared (FIR) thermal dust emission \citep[for an overview, see the `ring' class in][]{Cox2012}, suggesting that gas and dust seem to be mostly spatially aligned. Ultimately, when observed at higher spatial resolution and sensitivity with ALMA, one of the known detached shell targets, \object{R~Scl}, revealed a spiral pattern inside the spherically symmetric envelope \citep{Maercker2012,Maercker2016}; caused by a binary companion, the spiral windings allowed for an approximate continuous back tracing of the recent gas mass-loss evolution.

Despite the compelling observational evidence of a variable mass loss, even strong changes in mass-loss rate alone, however, are not expected to be sufficient to produce the geometrically thin and well-confined gas and dust structures, as presented in, for example, \citet{Kerschbaum2017}, \citet{Mecina2014}, and the literature referenced above. It is likely that a wind-wind interaction between outflows of differing velocities amplifies the initial enhancement in the radial density distribution \citep{Mattsson2007}. Nevertheless, in such a scenario, the mass-loss history of an AGB star can, in principle, be traced back a few thousand years for objects nearby (a few 100\,pc, that is), and provide constraints for stellar evolution models.

As an alternative to the wind-wind explanation, the interaction of the stellar outflow with the surrounding interstellar medium can, in principle, also explain the occurrence of detached shells \citep{Young1993a,Libert2007}. In this case, the shell forms when an increasing amount of matter is piled up as the wind progresses into the ISM.

In this paper, we present FIR Herschel/PACS observations and radiative transfer modelling of the circumstellar dust environment for two examples of detached shell objects, namely, \object{U~Hya} and \object{W~Ori}. This is done in a similar fashion as in a previous paper \citep{Mecina2014}, where the carbon stars S~Sct and RT~Cap were investigated. Besides a straightforward radiative transfer model with a parametrised density distribution, here, we also try to establish a more elaborate picture by means of stationary wind models for one of the targets.

\section{Observations}
The two targets presented in this paper are part of a much larger sample of objects that were observed in the course of the \textit{Mass loss of Evolved StarS} survey \citep[MESS,][]{Groenewegen2011}. Both stars had been previously studied, but only for U~Hya data showing extended circumstellar structures existed. 

\subsection{Sources}

\begin{table}
\centering
\caption{Source parameters.}
\label{tab:targets}
\begin{tabular}{lccccc}
\hline
\hline
\vspace{-9pt}\\
Target & Spec. type & $P$ & $D$ & $T_{\mathrm{eff}}$ & C/O \\
 & & [days] & [pc] & [K] & \\
\hline
\vspace{-8pt}\\
U~Hya & C6.5,3(N2)(Tc) & 450 & 208 & 2965 & 1.04 \\
W~Ori & C5,4(N5) & 212 & 377 & 2625 & 1.16 \\
\hline
\end{tabular}
\tablefoot{Variability and spectral data are taken from the GCVS \citep{Samus2017}, distances from Hipparcos parallaxes \citep{vanLeeuwen2007}. Effective temperatures are from \citet{Bergeat2001} and C/O from \citet{Lambert1986}.}
\end{table}

\subsubsection{U Hya}

The carbon star U~Hya is a semi-regular pulsator of type SRb with a main pulsation period of 450 days \citep{Samus2017}. It is one of the apparently brightest carbon stars in the sky \citep[V=4.82,][]{Ducati2002} at an estimated distance of $208 \pm 10\,$pc \citep{vanLeeuwen2007}. Recent Gaia data put it even closer at $172\,^{+17}_{-14}\,$pc \citep{GaiaCollaboration2018}. However, the nature of a typical AGB star -- namely, its large apparent diameter (up to the scale of the parallax itself), variability, and time-dependent surface features shifting the photocentre -- systematically hamper a more precise distance estimate via the parallax method \citep{Chiavassa2018}.

For the atmospheric C/O ratio, \cite{Lambert1986} found a low value of 1.04, suggesting that a few thermal pulses and subsequent dredge-ups have taken place. Evidence that this last such event may have occurred more recently (i.e. not more than a few $10^4$ years ago) is indicated by the detection of the unstable technetium isotope $^{99}$Tc \citep{Peery1971}. An overview of additional stellar parameters is given in Table~\ref{tab:targets}.
It has been known for some time that U~Hya is surrounded by an extended shell of cold dust -- following the initial observational evidence, in the form of FIR excess, which can be seen, for example, in the IRAS colour-colour diagram \citep{Veen1988}. Also, \cite{Waters1994} were able to confirm just resolved dust structures in highly processed IRAS imaging data. With the AKARI satellite, an improved view was later presented by \cite{Izumiura2011}. In that study, maps taken at 65, 90, 140, and $160\,\mathrm{\mu m}$ show a rather spherically symmetric structure that is detached from the central source.

Recently, using SCUBA-2 on the James Clerk Maxwell Telescope, \cite{Dharmawardena2018} detected extended sub-mm continuum emission at 850\,$\mathrm{\mu m}$ around U~Hya, which can be ascribed to a detached dust shell. There has been no detection thus far of a counterpart in molecular line emission, although observations in sub-mm CO transitions have been conducted by \cite{Olofsson1993a}.

\subsubsection{W Ori}

W~Ori is a variable carbon star of type SRb with a pulsation period of 212 days \citep{Samus2017} and \cite{Lambert1986} found an atmospheric C/O of 1.16. The distance for this object is quite uncertain, even in the realm of AGB stars. From reprocessed Hipparcos parallax measurements, \cite{vanLeeuwen2007} derived a value of $377_{-100}^{+211}$\,pc, whereas recent data from the Gaia DR2 yield a much larger distance of $1010_{-190}^{+320}$\,pc \citep{GaiaCollaboration2018}. In view of the already high luminosity derived for the Hipparcos distance (see Sect.~\ref{results}), the almost three times larger Gaia value requires a luminosity that is not consistent with a star in the AGB mass range.  Moreover, an estimate following the $P\mbox{-}L$ relation derived for SRb variables by \cite{Knapp2003}, gives a value of $290_{-60}^{+70}\,$pc, which is more in agreement with the Hipparcos parallax. Throughout this paper, we adopt the value from \citeauthor{vanLeeuwen2007} in our calculations.

Before Herschel, there was no observational evidence of extended circumstellar structures around W~Ori. \citet{Schoeier2001} only detected unresolved CO line emission, which they ascribed to a mass outflow at a rate of $7 \times 10^{-8}\,\mathrm{M_\odot/yr}$ with an expansion velocity of 11.0 km/s.

\subsection{Herschel/PACS observations}

The target stars were observed using \textit{Herschel/PACS} \citep{Poglitsch2010}, which  provides FIR imaging in three wavelength bands and spectroscopy. We used the $\mathrm{70\ \mu m}$ and $\mathrm{160\ \mu m}$ filters of the photometer array. The pixel scale of the detectors consisting of multiplexed bolometers is 3\farcs 2 and 6\farcs 4 for the short and long wavelength band, respectively. That sufficiently samples the typical telescopes' PSFs with a FWHM of 5\farcs 6 and 11\farcs 4, respectively.

U~Hya was observed on operation day (OD) 581 with observation IDs 1342212001 \& 1342212002. W Ori was observed on two occasions, the first time on OD 284, and again in a follow up campaign on OD 871 (observation IDs 1342190965 \& 1342190966 and 1342229983 \& 1342229984, respectively). All data were collected within the guaranteed time key programme MESS \citep{Groenewegen2011}.

The observations were conducted in scan map mode with medium scanning speed (20\arcsec/s) and a raw detector sampling rate of 10\,Hz (an averaging of four frames each had already been done on-board), where two scans with orthogonal scanning directions complement each other. This yields a highly noise-dominated stream of data, requiring rather extensive preprocessing of the individual bolometer pixel time lines, which was facilitated by using \textit{HIPE}\footnote{Herschel Interactive Processing Environment}. The processed time series were projected onto a grid with a spatial resolution of 1\arcsec\ for the $\mathrm{70\, \mu m}$ map and 2\arcsec\ for $\mathrm{160\, \mu m}$ map, respectively. Such an oversampling of the physical detector resolution by about a factor of 3 improves the rendering of the finest spatial structures. We considered a number of different mapping approaches and arrived at using \textit{JScanam}, the HIPE implementation of \textit{Scanamorphos} \citep{Roussel2013}, which, in our case, proved to offer the best compromise between smooth background rendering, good extraction of faint extended structures, and absence of (pre-)processing artefacts. An in-depth analysis of PACS bolometer array image data reconstruction techniques, including various mapping algorithms was conducted by \cite{Ottensamer2011} and \cite{Mecina2014a}.

\section{Modelling}

We first model the circumstellar dust envelope of both targets presented in this paper by means of radiative transfer, using \textit{More of DUSTY} \citep[MoD,][]{Groenewegen2012}, a minimisation wrapper for the 1D dust radiative transfer code \textit{DUSTY} \citep{Ivezic1999}. This approach was already taken for modelling other sources of the MESS sample \citep{Mecina2014}. In addition to this method, which is simply based on a parametrised dust density distribution, we also try to establish a more physical representation for the example of U~Hya with stationary wind models. In both cases, the spherical symmetry of the outflow justifies a 1D approximation of the envelope. In the following sections, the mass and the mass-loss rate quantities refer to the dust component only, except where explicitly stated otherwise.

\begin{table}
\centering
\caption{Photometric data used in the SED fitting.}
\label{tab:sed}
\begin{tabular}{rrr}
\hline
\hline
\vspace{-10pt}\\
Filter & Magnitude & Reference \\
\hline
\hline
\vspace{-8pt}\\
\multicolumn{3}{c}{U~Hya} \\
\hline
     V &     4.820 & \citet{Morel1978} \\
     R &     3.050 & \citet{Morel1978} \\
     I &     1.780 & \citet{Morel1978} \\
     J &     0.820 & \citet{Morel1978} \\
     K &    -0.750 & \citet{Morel1978} \\
     L &    -1.180 & \citet{Morel1978} \\
2massJ &     0.803 & \citet{Cutri2003} \\
2massH &    -0.254 & \citet{Cutri2003} \\
2massK &    -0.716 & \citet{Cutri2003} \\
AkaS9W &    -1.614 & \citet{Ishihara2010} \\
AkL18W &    -2.152 & \citet{Ishihara2010} \\
IRAS12 &    -1.753 & \citet{Beichman1988} \\
IRAS25 &    -2.206 & \citet{Beichman1988} \\
PACS70 & 35767.000 & this work \\
PACS160& 16200.000 & this work \\
\hline
\vspace{-8pt}\\
\multicolumn{3}{c}{W~Ori} \\
\hline
  V &    6.170 & \citet{Morel1978} \\             
  R &    3.850 & \citet{Morel1978} \\             
  I &    2.370 & \citet{Morel1978} \\             
  J &    1.510 & \citet{Epchtein1990} \\             
  H &    0.320 & \citet{Epchtein1990} \\             
  K &   -0.340 & \citet{Epchtein1990} \\             
  L &   -1.040 & \citet{Epchtein1990} \\             
  M &   -0.530 & \citet{Epchtein1990} \\                       
AkaS9W &   -1.447 & \citet{Ishihara2010} \\          
AkL18W &   -1.950 & \citet{Ishihara2010} \\          
AkarWS &   -2.661 & \citet{Yamamura2010} \\          
AkarWL &   -2.854 & \citet{Yamamura2010} \\  
IRAS12 &   -1.630 & \citet{Beichman1988} \\          
IRAS25 &   -1.841 & \citet{Beichman1988} \\          
IRAS60 &   -2.391 & \citet{Beichman1988} \\        
PACS70 & 9593.000 & this work \\          
PACS160& 2227.000 & this work \\          
\hline
\hline
\end{tabular}
\tablefoot{For the PACS photometry, the flux in units of mJy is given instead of the magnitude.}
\end{table}

\subsection{MoD dust radiative transfer}
\label{modmodelling}

For the MoD approach, we follow the same procedure as described in \cite{Mecina2014}, that is, for the dust density, $\rho,$ we adopt a piecewise power law distribution. The innermost part describes the present-day wind, assuming a continuous smooth outflow starting at the condensation radius, where the fixed dust temperature, $T_\mathrm{c},$ is reached. In the case of constant mass loss, $\rho,$ simply drops proportionally to $r^{-2}$ due to geometrical rarefaction with radius, $r$, but we allow for deviating density profiles in the sense of a variable exponent of the power law. At a certain (free) radius, the inner wind region is followed by a shell of variable thickness $\delta r$ and density scaled by a variable factor of $s_1$. Within this shell, $\rho$ is kept proportional to $r^{-2}$. Outside the shell, the density is then fixed to negligibly low values (typically 0.1\% of the detached shell density at the outer border). This is representative of the ISM and pre-high-mass-loss conditions and has no influence on our results. The radiation source in the centre is selected from the grid of COMARCS model atmospheres from \cite{Aringer2009}, based on the stellar parameters given in Table~\ref{tab:targets} and reasonable assumptions concerning stellar mass and surface gravity. Since all computed DUSTY models are scale-free, initially we only need the relative shape of the input spectrum, that is, we are not interested in the actual luminosity of the COMARCS model. Instead, it serves as a free parameter that is determined within MoD by fitting the SED to photometric data shorter than $12\,\mathrm{\mu m}$, thus neglecting the FIR detached shell emission and only taking the warm dust in the present-day mass loss into account. The obtained luminosity value is then adopted and kept fixed for the full modelling of the envelope. All photometric data points are de-reddened for interstellar extinction according to the target's galactic position. For this we use the same method as \cite{Mecina2014} that is described in \cite{Groenewegen2008}.

A key ingredient in the radiative transfer models is the dust opacities. We calculate these using Mie theory \citep[BHCOAT,][]{Bohren1983}, using optical constants from \cite{Rouleau1991} for amC and \cite{Pegourie1988} for SiC. The two species are not treated separately in the radiative transfer but, rather, approximated by a single opacity table with weighted contributions from the respective substances. This is justified in view of the similar overall opacity distributions and condensation temperatures. Instead of solid spherical grains, we adopt a distribution of hollow spheres \citep[DHS,][]{Min2003} with a size of 0.15\,$\mathrm{\mu m}$ and a varying vacuum fraction of up to 0.7, following the approach of \citet{Groenewegen2012}. These are values typically used for modelling porous circumstellar dust. Changing either the optical constants by choosing another source of laboratory measurements or adopting different grain shapes and size distributions can alter the resulting models significantly, as seen in a recent study by \cite{Brunner2018}. The findings therein, exemplarily given for the carbon star R~Scl, are also applicable to the calculations in this paper. In a statistical study of carbon stars in the Small Magellanic Cloud, \citet{Nanni2016} have tried to identify the most suitable optical constants for amorphous carbon and the typical grain size to which dust particles grow in that environment. Their findings, however, have to be taken with caution with regard to the sources in this paper since different stellar populations are considered and another grain geometry (solid spheres) was adopted.

The model output consists of photometric fluxes, a spectrum, and radial intensity profiles, which can strongly constrain the spatial scales. We find the model with the best representation of the observations by comparing the output to photometric data from the literature (Table~\ref{tab:sed}) and the FIR fluxes obtained in this paper (Table~\ref{tab:pacsphoto}), along with mid IR spectra and averaged brightness profiles from the PACS maps. The goodness of the fit of a model is evaluated by calculating the $\chi^2$, where
\begin{equation}
\chi^2=\sum \limits_{i=1}^{n}(m_\mathrm{obs}(i)-m_\mathrm{pred}(i))^2/\sigma^2_\mathrm{m_{obs}(i)}.
\label{eq:chi2}
\end{equation}
Further details about the model fitting can be found in \cite{Groenewegen2012}.

\subsection{Stationary wind models}
\label{windmodels}

We want to compare the results of our radiative transfer calculations (based on a parametrised density profile) with another modelling approach, where the dust distribution is predicted by the model. For this, we use the code from \cite{Ferrarotti2006}, which describes dust formation in a stationary stellar outflow and the subsequent dust-driven wind. This method has the advantage that, for given elemental abundances and dust opacity data, only realistic combinations of the inner radiation source, mass-loss rate, and dust distribution are possible. In addition, the gas-to-dust ratio can be obtained from the calculations. A similar model approach was already used for carbon stars by \citet{Nanni2016,Nanni2019}.

For the modelling presented here, we applied an enhanced version of the wind code.\footnote{http://www.ita.uni-heidelberg.de/\textasciitilde gail/agbdust/agbdust.html}
The main features were already described in \citet{Ferrarotti2006}. On top of these, a few improvements were implemented (e.g. additional dust species were included, the effect of particle drift was considered, etc.). The most important change was the additional implementation of a radiative transfer module which self-consistently calculates temperatures for each dust species by solving the radiative transfer problem in the dust shell and iterating to flux constancy. In this way, the coupling between the dust condensation in the outflow and the radiation field is treated in a more consistent way. This enhanced version of the wind code was already applied in \citet{Mayer2013} and \citet{Nowotny2015}.

\begin{figure}[ht]
\includegraphics[width=\hsize]{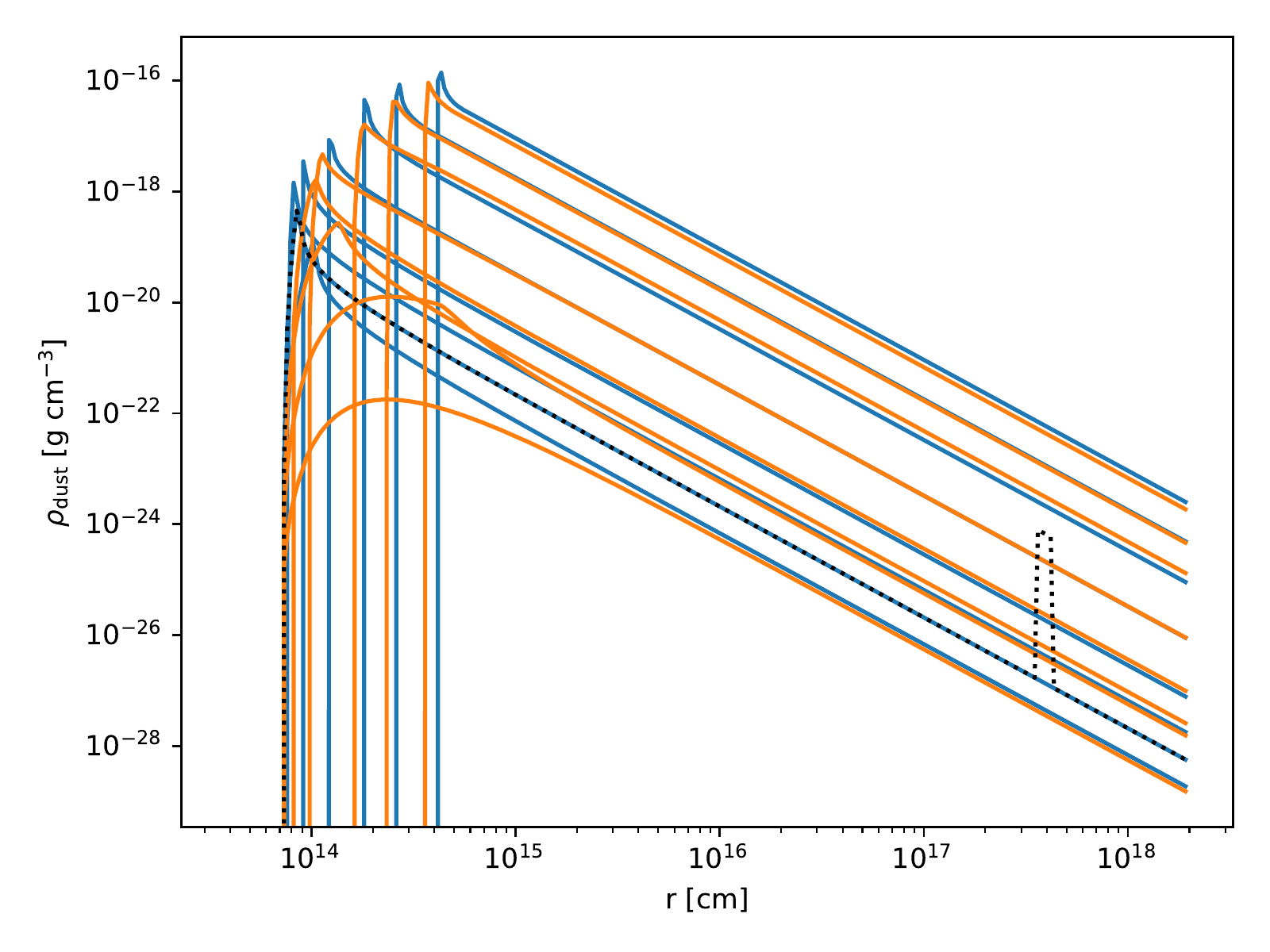}
\caption{Dust-density profiles of the baseline stationary wind models with varying MLRs. Blue and orange indicate low (0.5\,km/s) and high (1.5\,km/s) initial outflow velocity, respectively. The dotted line is a model with detached shell, where the density is scaled by a factor of 500.}
\label{fig:density_profs}
\end{figure}

The main free parameters of the stationary wind models are the present-day total mass-loss rate (pdMLR) and the initial outflow velocity ($u_{ini}$). We made a small grid covering two typical values of $u_{ini}$ (0.5 and 1.5\,km/s) and total mass-loss rates between $3\times 10^{-8}$ and $10^{-4}\,\mathrm{M_\odot/yr}$ (see Table~\ref{tab:coma_grid}). An overview of the resulting dust density distributions is given in Fig.~\ref{fig:density_profs}.

Concerning the selection of the central star, we follow the approach of our MoD fits, adopting a hydrostatic atmosphere model with fixed parameters. This was taken from the grid of \citet{Aringer2016,Aringer2019}. Based on the data given in Table~\ref{tab:targets}, the effective temperature was set to 3000\,K, which is consistent with the choice for the MoD fits. The $\rm log(g\,[cm/s^2])$ value of $-0.14$ and the mass of 2\,M$_\odot$ were selected to be in agreement with the derived luminosity of the star. It should be noted that the choice of these two parameters has no significant impact on the overall energy distribution of the central source. In contrast to the DUSTY models used for the MoD fits, the luminosity is a crucial quantity in the wind models. For the abundances we assumed a solar mixture \citep{Caffau2009,Caffau2009a}, except for carbon, which was increased to obtain a C/O of 2. Such a high value is in contradiction with the low C/O values in Table~\ref{tab:targets} and in the central sources of the MoD fits. Nevertheless, it is a necessary condition for an effective dust production in the wind. This discrepancy will be discussed in Sect.~\ref{discussion} in more detail. Finally, we want to note that the elemental abundances adopted for the central atmosphere and the dust formation in the stellar wind are identical.

Of course, in such an outflow, structures like the observed detached shells, which have been claimed to be the consequence of highly variable mass loss, cannot occur as per the relevant definition. Therefore, we need to artificially introduce a non-stationary behaviour by increasing the densities in the respective regions of the model structure of the wind. In our case, we simply multiplied the density profile by a constant factor at the radial points that were derived in the MoD fitting and which are well-constrained by the PACS maps. The density contrast between the inner wind structure and the detached shell is treated as a free parameter. We test values between 10 and 10000 for each of the models in the grid, except for baseline models with a present-day gas MLR of $10^{-5}\,\mathrm{M_\odot/yr}$ or higher. In these cases, the resulting mass-loss rates corresponding to the detached shells would have unrealistically large values (see the overview in Table~\ref{tab:coma_grid}).

Based on the model structures, we computed synthetic opacity sampling (OS) spectra with a resolution of R~=~10000 and the corresponding frequency-dependent radial intensity profiles using the COMA code \citep{Aringer2009,Aringer2016}. Based on the results of these calculations, which cover the range between 0.335 and 200~$\mu$m, we could determine the photometric magnitudes, low-resolution spectra, and flux distributions in the Herschel/PACS images. The COMA results were obtained with the same elemental abundances and opacity data for atoms, molecules, and dust as the central atmospheric and the circumstellar structures. However, we neglected the line absorption due to gas in the wind regions because it has only a minor impact on the overall energy distributions and the assumption of LTE and chemical equilibrium in the COMA computations may become quite problematic in such dynamic and low-density environments.

The stationary wind models include amC and SiC. However, in contrast to the MoD approach, these two species are always treated as two separate components with their own radial temperature, density, and grain size distribution. In addition, their growth along the outflows is considered. Instead of a fixed DHS, as used in the MoD fits, the geometric shapes are assumed to be spheres.

\begin{figure}[ht]
\includegraphics[width=\hsize]{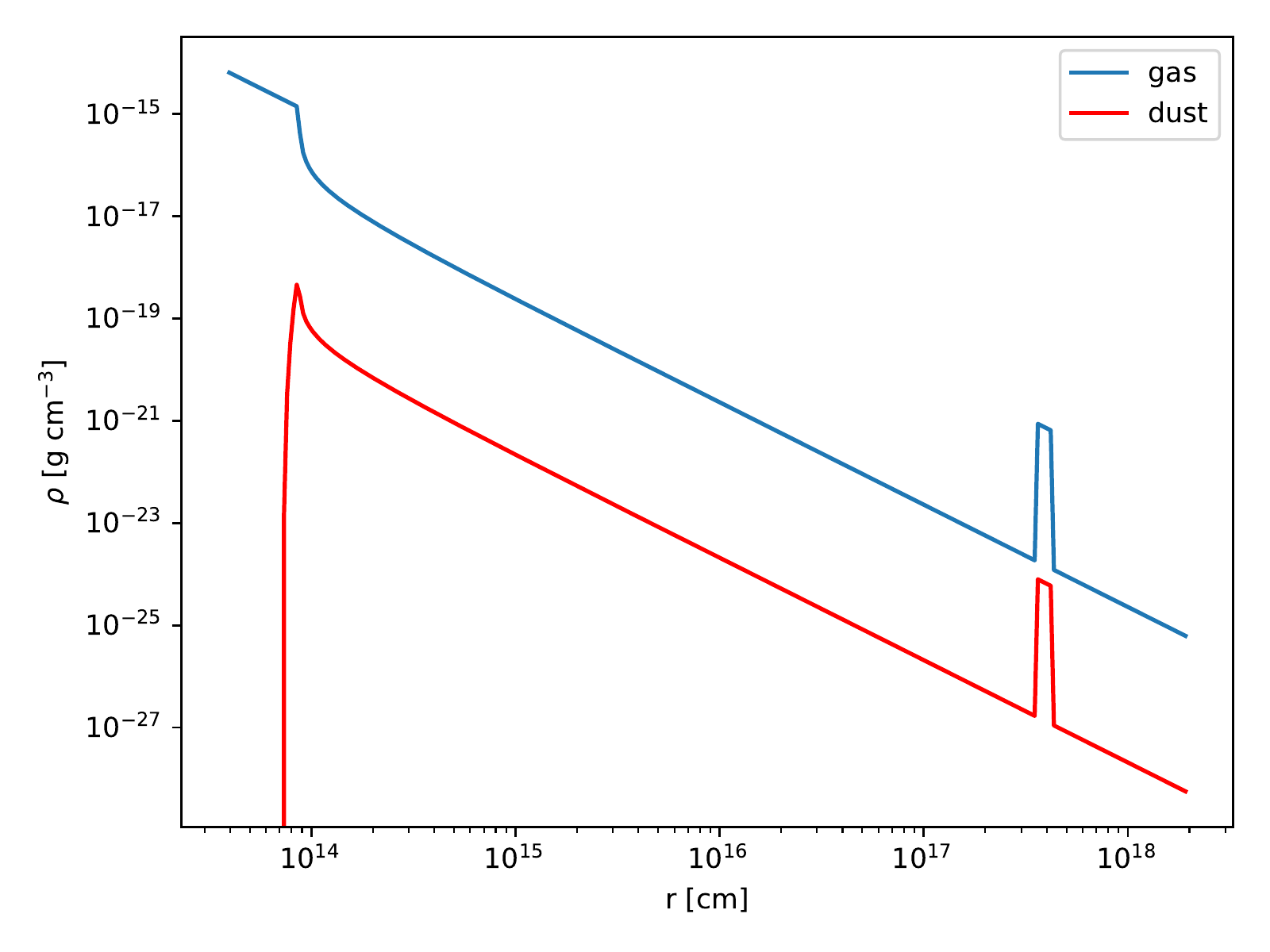}
\caption{Comparison of the gas and amC dust-density distribution for a model with a total MLR of $10^{-7}\,\mathrm{M_\odot/yr}$ and the detached shell density enhanced 500-fold.}
\label{fig:gas_dust_profs}
\end{figure}

With the  above-described method, we computed a relatively coarse grid of models. In addition to the initial and present-day total mass-loss rate (pdMLR) we vary the amplitude $\Delta\rho$ and width $\Delta r_\rho$ of the enhancement in the density structure representing the detached shell. The covered parameter range is given in Table~\ref{tab:coma_grid} and an example of a resulting gas and dust density profile is shown in Fig.~\ref{fig:gas_dust_profs}. Due to the substantial effort of computing such a grid and the very uncertain distance and luminosity of W~Ori, here, we only apply the models to the example of U~Hya.

\begin{table}
\centering
\caption{Overview of the stationary wind model grid. Initial outflow velocity $u_\mathrm{ini}$ and present-day total MLR are the input parameters. Columns amC and SiC give the resulting dust MLRs for the respective species. $\Delta\rho$ lists the density scaling ranges for the detached shells applied to the respective models. Each density enhancement was scaled in width $\Delta r_\rho=(4-8)\times10^{16}$\,cm.}
\label{tab:coma_grid}
\begin{tabular}{c@{}cccl}
\hline 
\hline 

 $u_\mathrm{ini}$ [km/s] & \multicolumn{3}{c}{pd MLR [M$_\odot$/yr]} & $\Delta\rho$\\
 & total & amC & SiC \\ 
\hline 

0.5 & 3E-08 & 2.72E-11 & 6.54E-15 & 10-10000 \\
1.5 & 3E-08 & 1.78E-12 & 1.23E-15 & 10-10000 \\
0.5 & 1E-07 & 9.80E-11 & 6.18E-14 & 10-5000 \\
1.5 & 1E-07 & 7.68E-11 & 2.68E-14 & 10-5000 \\
0.5 & 3E-07 & 4.25E-10 & 2.16E-12 & 10-2500 \\
1.5 & 3E-07 & 3.28E-10 & 2.07E-13 & 10-2500 \\
0.5 & 1E-06 & 1.61E-09 & 1.68E-10 & 10-1000 \\
1.5 & 1E-06 & 1.52E-09 & 4.67E-12 & 10-1000 \\
0.5 & 3E-06 & 8.03E-09 & 1.30E-09 & 10-500 \\
1.5 & 3E-06 & 6.63E-09 & 1.87E-10 & 10-500 \\
0.5 & 1E-05 & 4.98E-08 & 4.55E-09 & - \\
1.5 & 1E-05 & 4.57E-08 & 4.33E-09 & - \\
0.5 & 3E-05 & 1.53E-07 & 1.36E-08 & - \\
1.5 & 3E-05 & 1.53E-07 & 1.36E-08 & - \\
0.5 & 1E-04 & 5.12E-07 & 4.56E-08 & - \\
1.5 & 1E-04 & 5.12E-07 & 4.55E-08 & - \\

\hline 
\hline 
\end{tabular} 
\end{table}

\section{Results}
\label{results}
We find morphologically very similar structures around the two target stars. While the one around U~Hya was already known, the detached shell around W~Ori is a new discovery. In the following we present the derived properties of the respective dust envelopes.

\subsection{Morphology}

\subsubsection{U Hya}
\label{uhya_morph}

\begin{figure*}[h]$
\begin{array}{cc}
\includegraphics[width=0.45\hsize]{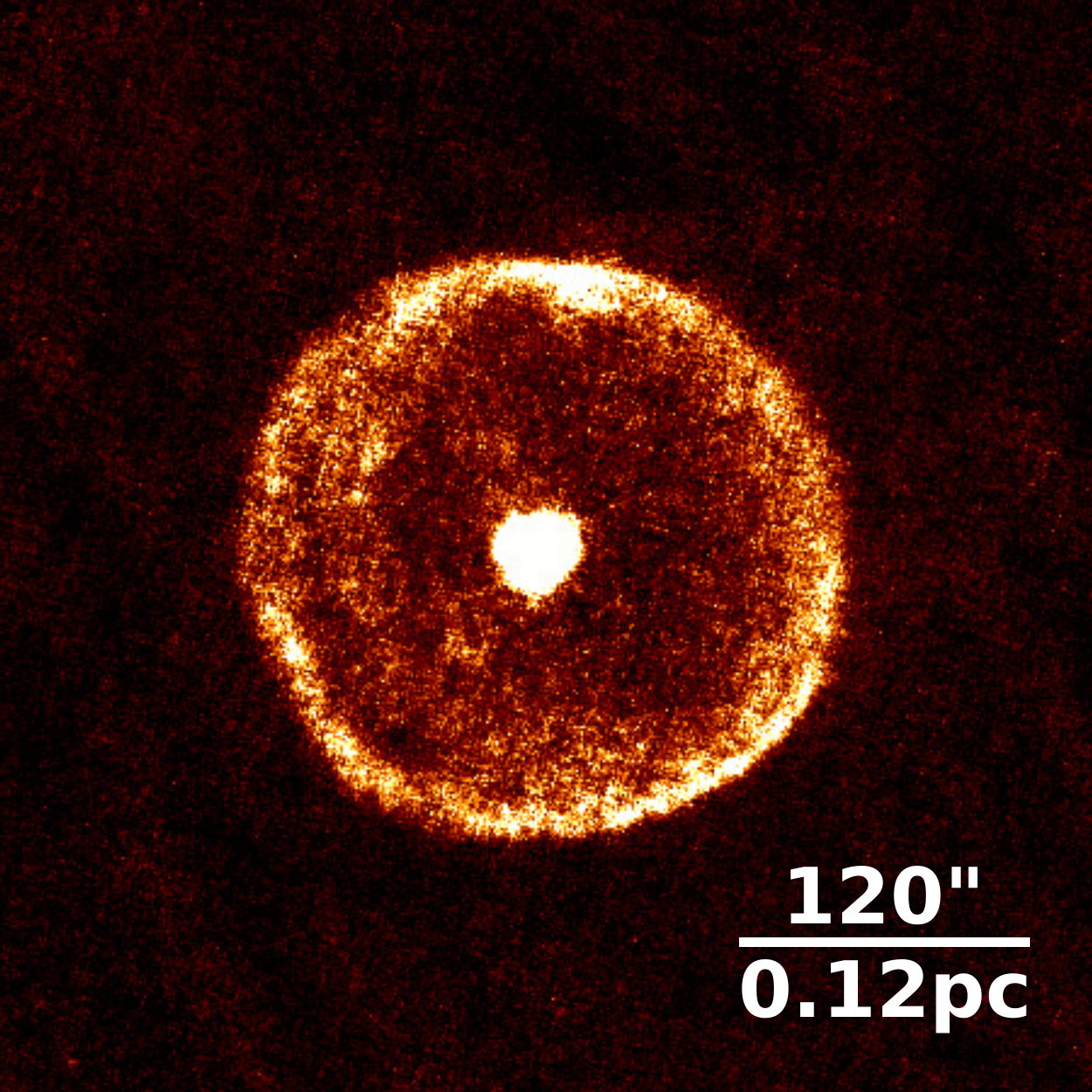} &
\includegraphics[width=0.45\hsize]{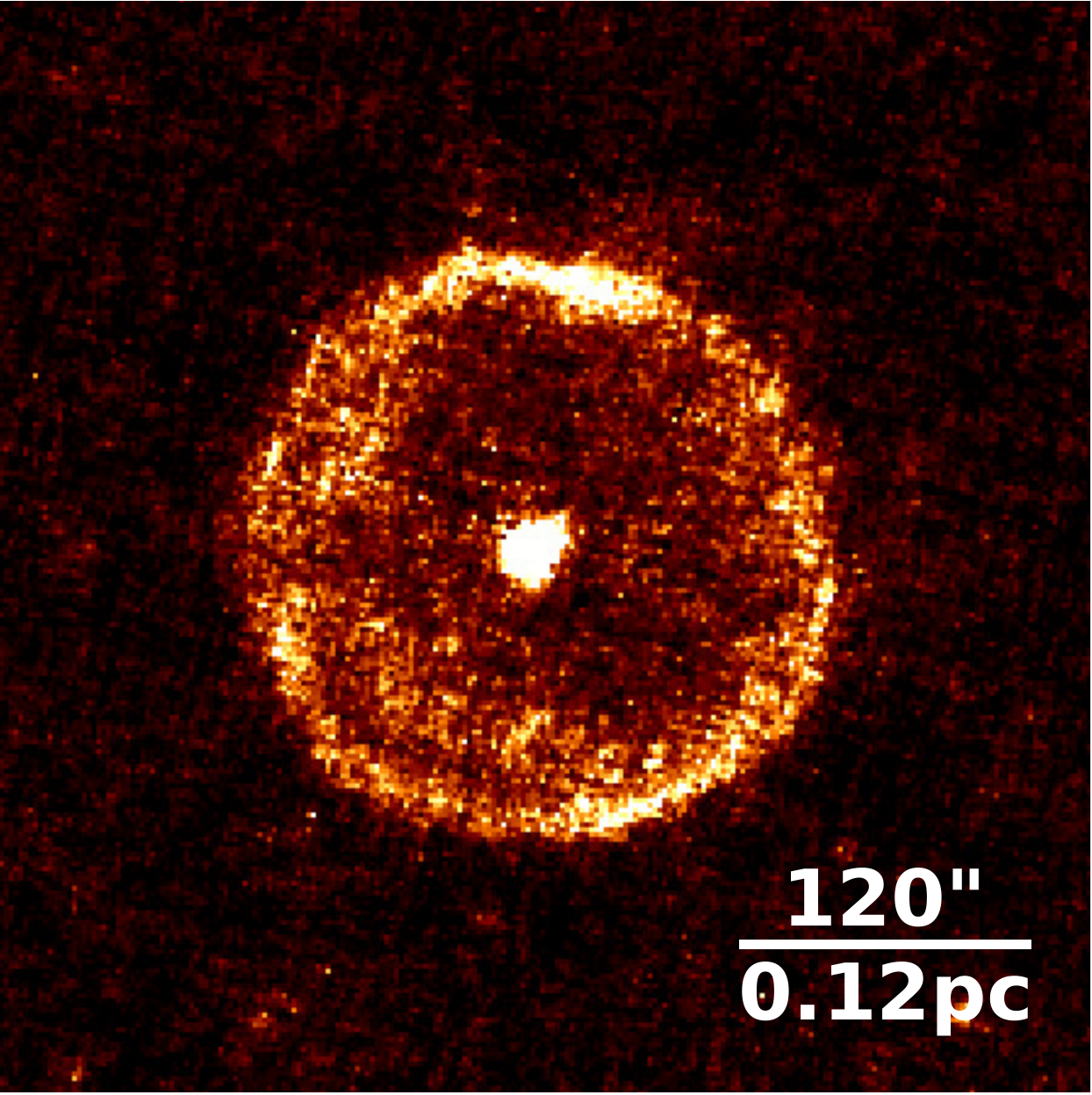}
\end{array}$
\caption{PACS maps of U~Hya at 70\,$\mu$m (left) and 160\,$\mu$m (right). The spatial resolution (FWHM) is 5\farcs 8 and 11\farcs 5, respectively.}
\label{fig:uhya_pacs}
\end{figure*}

\begin{figure*}[h]
\includegraphics[width=\hsize]{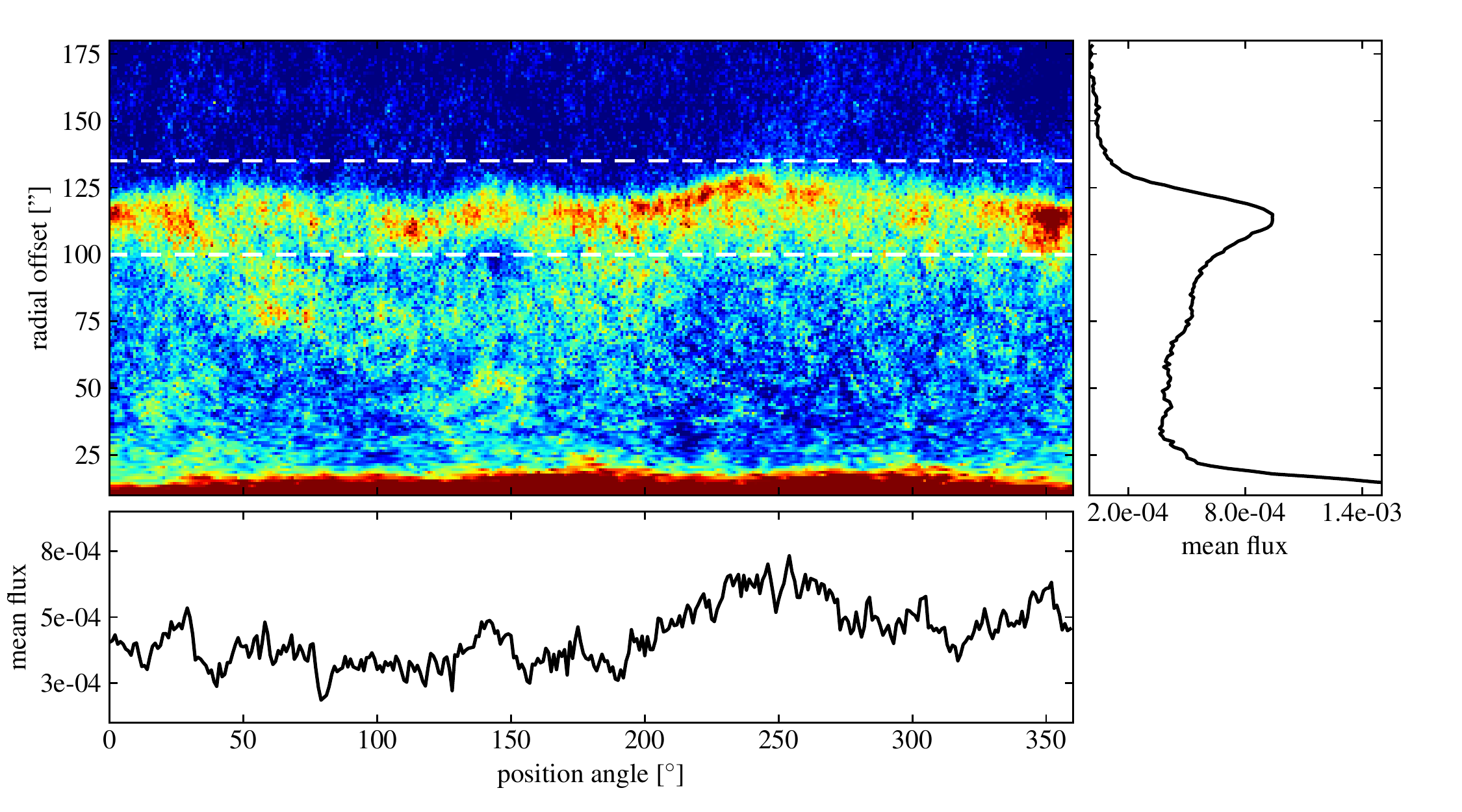}
\caption{Representation of U~Hya 70\,$\mu$m data in polar coordinates, centred on the star, with an azimuthally averaged radial profile (right). The radial average of the shell (bottom profile) was taken between 100\arcsec and 135\arcsec (dashed lines). Mean flux is given in Jy arcsec$^{-2}$.}
\label{fig:uhya_polar}
\end{figure*}

\begin{figure}
\includegraphics[width=\hsize]{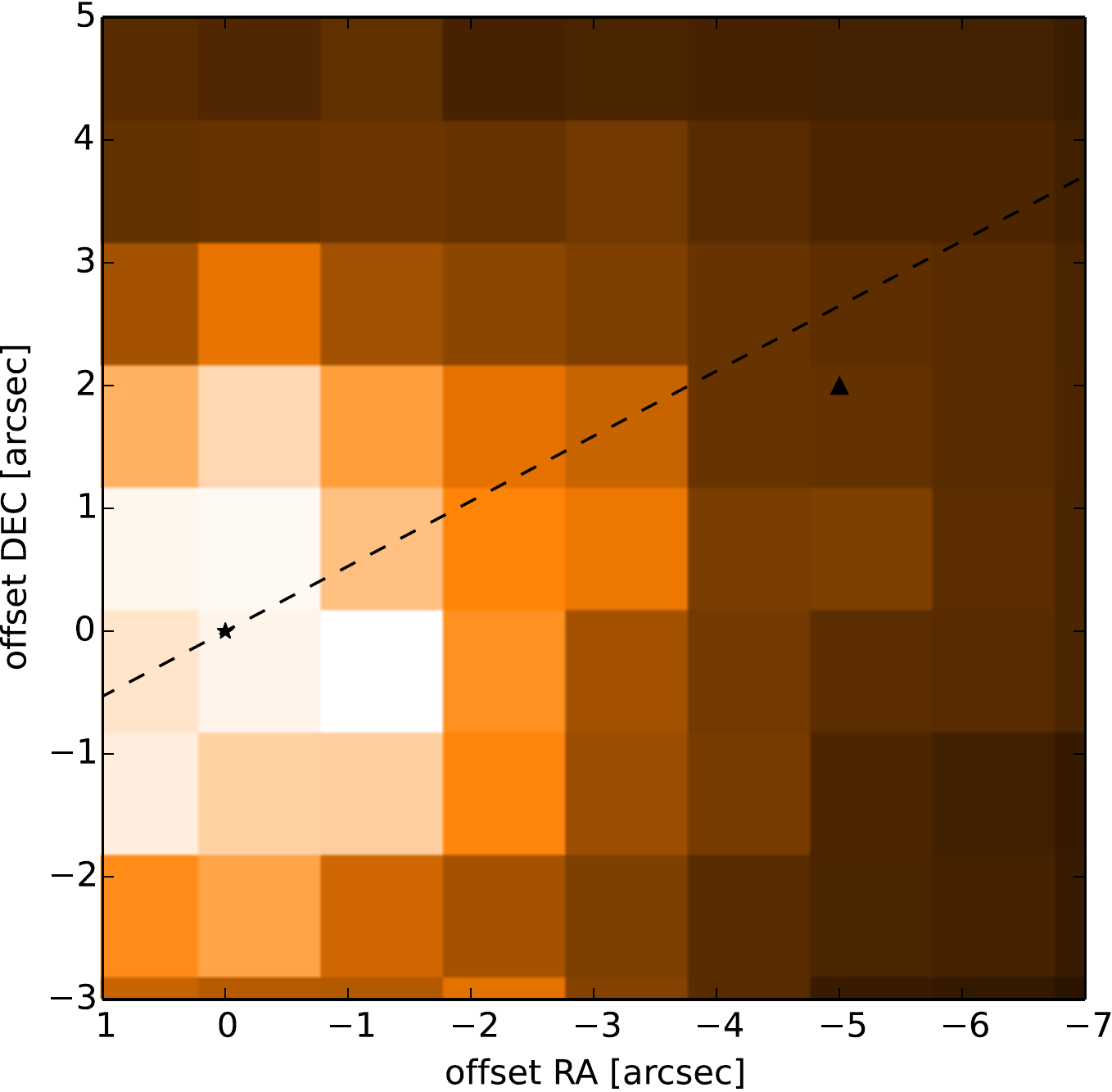}
\caption{Stellar position of U~Hya (derived from the 70\,$\mu$m PACS map) is offset from the shell centre (indicated by the triangle) by roughly 5\arcsec with a position angle of $\sim 110^\circ$. The dashed line represents the path of the space motion.}
\label{fig:uhya_astro}
\end{figure}

In both the blue and red Herschel/PACS filters, we observe thermal dust emission from a spherically symmetric shell detached from the central star (shown in Fig.~\ref{fig:uhya_pacs}), giving the impression of a ring-like structure. The intensity peaks at a radius of 114\arcsec, which, adopting a distance of 208\,pc, is equivalent to an absolute scale of 0.12\,pc. The emission is not evenly distributed along the circumference, but is stronger in the northern and south-western parts, congruent in the 70 and $\mathrm{160\,\mu m}$ maps. While we see patchy features in the shell, most likely we do not resolve its width, which puts an upper limit of roughly 1200\,AU on the radial thickness. Inside the detached shell in the south-east direction, we further detect diffuse, arc-shaped emission in both bands. The asymmetries and the clumped distribution of the emission become even more apparent when the PACS map is projected to polar coordinates, as can be seen in Fig.~\ref{fig:uhya_polar}.

In order to determine the geometric centre of the shell, we fit the (higher resolved) $\mathrm{70\ \mu m}$ emission with a Gaussian ring, that is, a ring with a radially Gaussian brightness distribution with the maximum at the shell radius. We find that the calculated position is offset relative to the stellar centroid by about $-5\arcsec$ in right ascension and $2\arcsec$ in declination. The results are displayed in Fig.~\ref{fig:uhya_astro}, where we mark the coordinates of the stellar source and the centre of the shell in the $\mathrm{70\ \mu m}$ PACS image. The shift is also optimally visible in polar coordinates (Fig.~\ref{fig:uhya_polar}), where the shell emission clearly deviates from a straight horizontal line, which would be expected for a perfectly symmetric geometry. The displacement can, in principle, be explained by the star's relatively high space velocity (71\,km/s) when the shell is influenced by the ISM headwind. That scenario is supported by the good alignment of the space motion vector (P.A. $=118^\circ$) with the orientation of the star-shell offset (P.A. $\approx110^\circ$). The slight deformation, or flattening, of the shell in the same direction would also support this interpretation. In Table~\ref{tab:kinematics}, we give a summary of the target's kinematic data, which we calculated following the method presented in \cite{Johnson1987}. 

\begin{table}
\centering
\caption{Aperture photometry from PACS maps. The total flux error is typically 5\% for the blue and 10\% for the red channel. The rightmost column gives the radii of the aperture where the detached shell fluxes were measured.}
\label{tab:pacsphoto}
\begin{tabular}{crrrrr}
\hline
\hline
 & \multicolumn{2}{c}{$70\,\mu$m} & \multicolumn{2}{c}{$160\,\mu$m} &\\
 & total & shell & total & shell & shell$_\mathrm{in}$-shell$_\mathrm{out}$\\
 & [Jy] & [Jy] & [Jy] & [Jy] & [$\arcsec$]\\
\hline
U~Hya & 37.8 & 15.7 & 16.6 & 7.8 & 100-135\\
W~Ori & 10.2 & 1.3 & 2.6 & 0.6 & 70-110\\
\hline
\end{tabular} 
\end{table}

\begin{table}
\centering
\caption{Summary of the space motion parameters for the two target stars, given in the heliocentric reference system (H) and corrected for solar motion in the local standard of rest (G). $\mu$ and PA give the absolute value and position angle of the proper motion vector, respectively. $v_\mathrm{r}$ is the radial velocity of the target. The derived space motion vector is defined by its absolute velocity $v_\mathrm{s}$ and its inclination with respect to the plane of sky $\theta$ (negative values indicate movement towards the observer).}
\label{tab:kinematics}
\begin{tabular}{lccccc}
\hline
\hline
\vspace{-9pt}\\
&$\mu$ [mas\,yr$^{-1}$]&PA&$v_\mathrm{r}$ [km\,s$^{-1}$]&$v_\mathrm{s}$ [km\,s$^{-1}$]&$\theta$\\
\hline
\vspace{-8pt}\\
\multicolumn{6}{l}{U~Hya} \\
H& 56.8 & $132^\circ$ & -25.8 & $61.8\pm2.5$ & $-25^\circ$\\
G& 63.9 & $118^\circ$ & -33.3 & $71.4\pm2.5$ & $-28^\circ$\\
\hline
\vspace{-8pt}\\
\multicolumn{6}{l}{W~Ori} \\
H& 7.6 & $101^\circ$ & $16.5$ & $21.4\pm2.8$ & $50^\circ$\\
G& 6.5 & $55^\circ$ & $2.6$ & $11.8\pm3.1$ & $13^\circ$ \\
\hline
\end{tabular}
\tablefoot{The space motion is calculated based on proper motion and $v_\mathrm{r}$ data from \citet{vanLeeuwen2007}.}
\end{table}

\subsubsection{W Ori}

\begin{figure*}$
\begin{array}{cc}
\includegraphics[width=0.45\hsize]{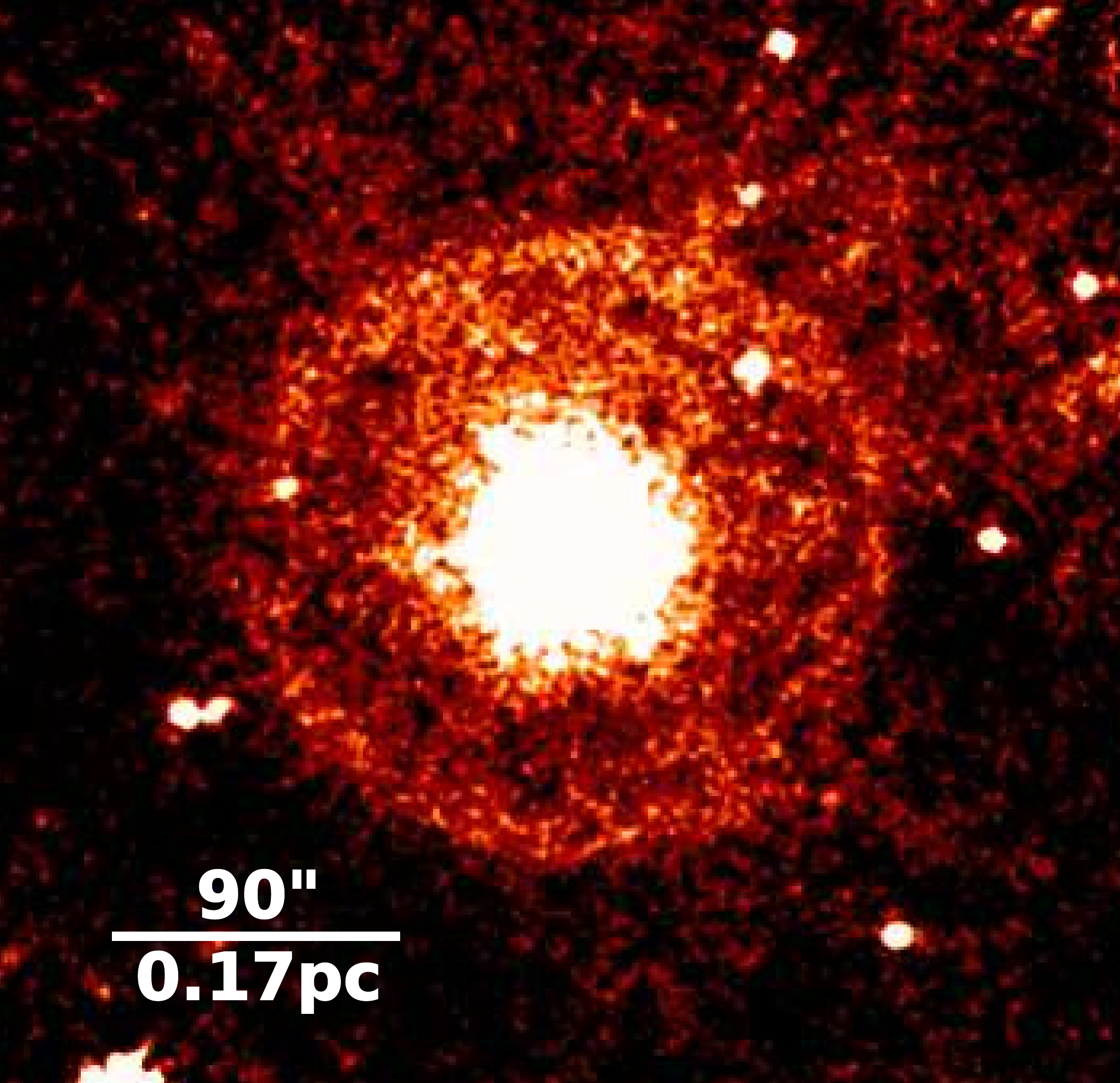} &
\includegraphics[width=0.45\hsize]{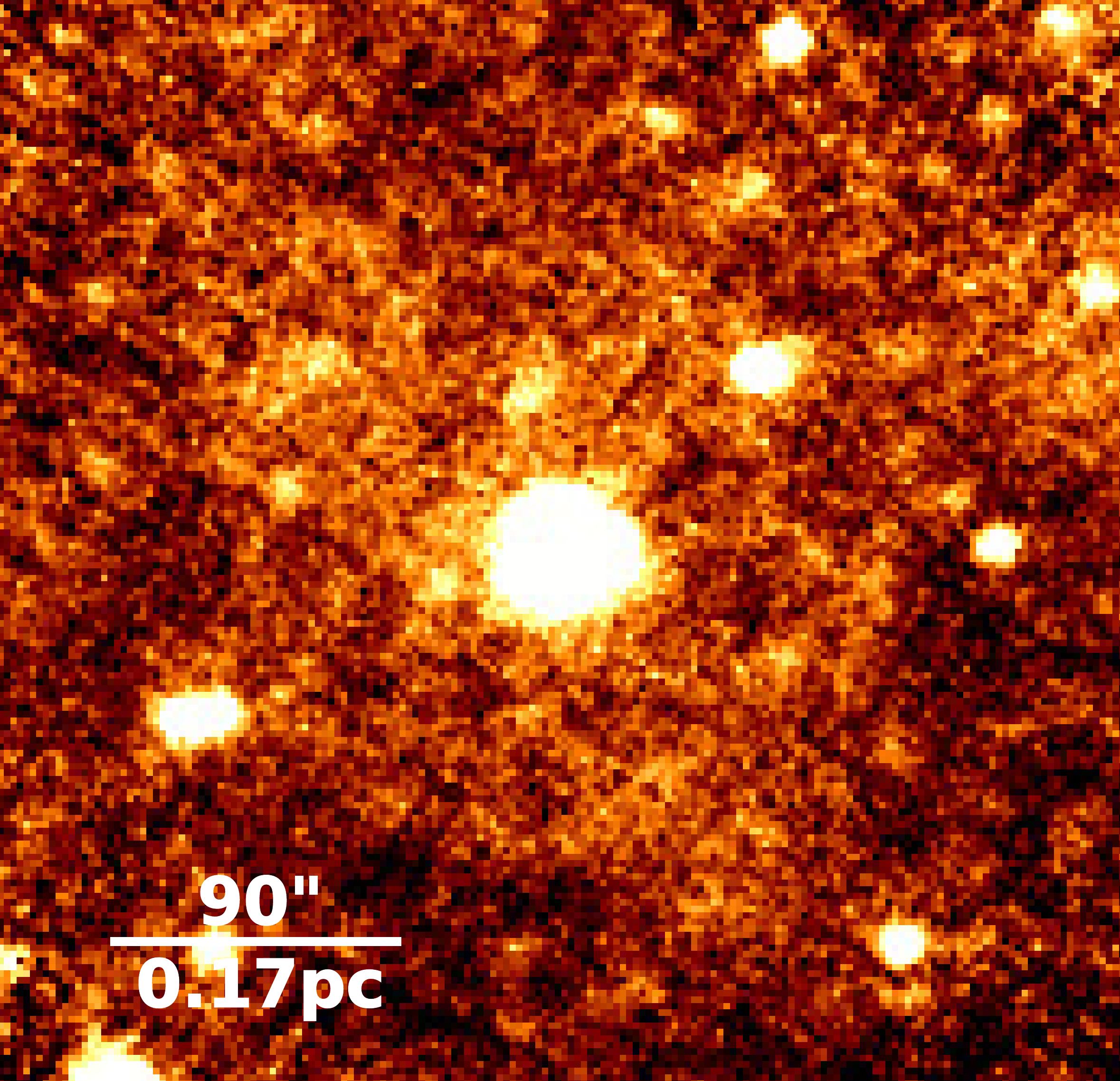}
\end{array}$
\caption{PACS maps of W~Ori at 70\,$\mu$m (left) and 160\,$\mu$m (right). The spatial resolution (FWHM) is 5\farcs 8 and 11\farcs 5, respectively.}
\label{fig:wori_pacs}
\end{figure*}

Both PACS maps, displayed in Fig.~\ref{fig:wori_pacs}, show a weak, spherically symmetric detached circumstellar shell. The dust emission peaks at $\sim92\arcsec$ from the central source, corresponding to a linear extent of 0.17\,pc. Keeping in mind the large uncertainty of the distance estimate, this value could be scaled up by a factor of almost 3. Just as for U~Hya, the shell width in the images is essentially determined by the instrument PSF. We can thus only give an upper limit of 2100\,AU for the thickness, however, this value could be higher as well, given the potentially larger distance.  The $\mathrm{70\ \mu m}$ emission appears to be rather homogeneously distributed azimuthally in the detached shell, however, the low flux levels hamper the identification of potential clumpy structures, as they appear around U~Hya. Moreover, the shell is contaminated by several background sources. The situation is even worse in the $\mathrm{160\ \mu m}$ band, where the shell is barely distinguishable from the diffuse galactic background emission and bright extragalactic sources significantly contribute to the flux measured within the circumstellar envelope. The latter considerably increases the uncertainty of the aperture photometry at that wavelength. The obtained values are given in Table~\ref{tab:pacsphoto}. 

Furthermore, we find the shell's centre to be well-aligned with the stellar position within the margins of error. Considering the low space velocity of the star (11.8\,km/s, see Table~\ref{tab:kinematics}), a substantial displacement of the shell due to ISM interaction is not expected.

\subsection{MoD models}

\subsubsection{U Hya}

\begin{figure}
\includegraphics[width=0.9\hsize]{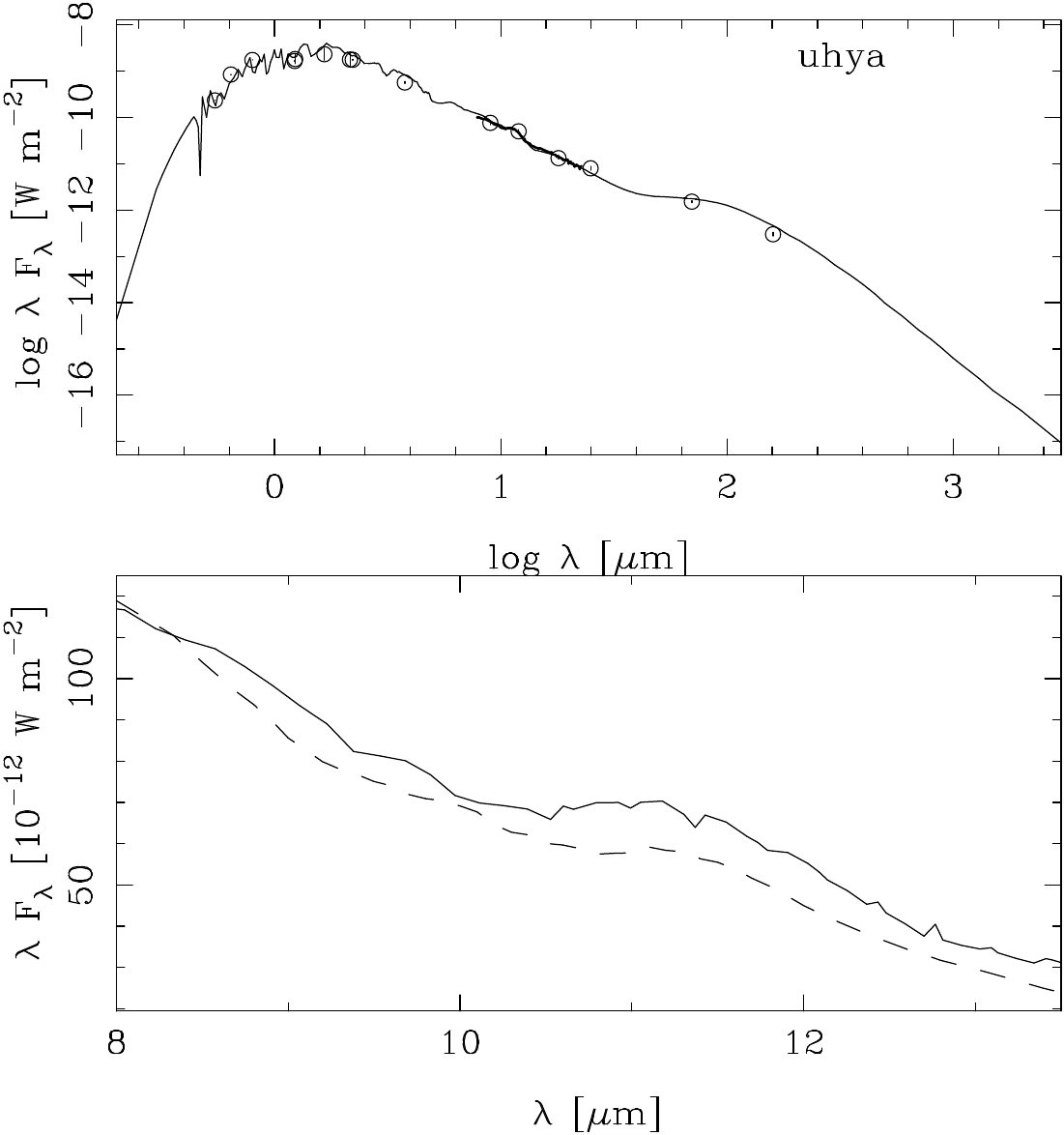}
\caption{Top: SED of the U~Hya MoD model fit compared to the photometric data (for references, see Table\,\ref{tab:sed}) and part of an IRAS LRS spectrum was also taken into account. Bottom: Comparison between the LRS (solid line) and the model spectrum around the SiC feature.}
\label{fig:uhya_sed}
\end{figure}

As discussed in Sect.~\ref{modmodelling}, we adopt an atmosphere model with $T_\mathrm{eff}=3000\,\mathrm{K}$ and a C/O of 1.05 as the central source. For the dust a mixture of 90\% amorphous carbon and 10\% SiC is used because the IRAS LRS spectrum \citep{Volk1989} shows the corresponding emission feature at $11.2\,\mathrm{\mu m}$. This practice is in agreement with previous works, for example, by \cite{Groenewegen2012a}. The condensation temperature is fixed at 1000\,K, a reasonable value for both carbon and SiC. As can be seen in Fig.~\ref{fig:uhya_sed}, the best-fit model SED is able to reproduce the photometric observations across all wavelengths. The modelled SiC spectral feature, however, is not as strong as in the LRS. Because a further increase of the emission in this feature would require an unreasonably high abundance of SiC in the dust mixture, it is probable that other factors such as molecular bands contribute to the observed shape. Also, a higher condensation temperature of SiC would result in a more prominent feature, but due to the limitation to a single dust opacity table, this cannot be accounted for in the MoD models.

\begin{figure}
\includegraphics[width=\hsize]{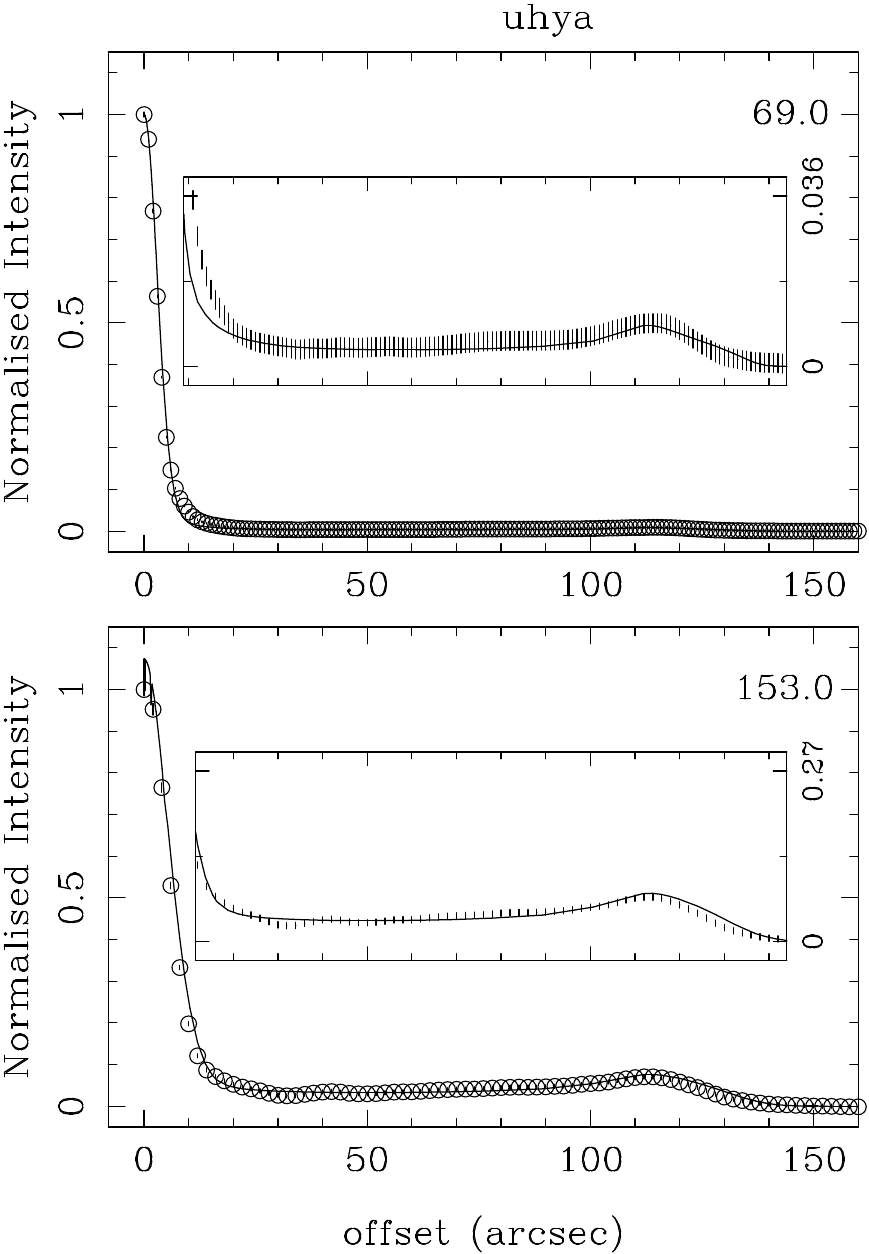}
\caption{MoD model fit and azimuthally averaged radial profiles of the PACS observations of U~Hya at 70\,$\mu$m (top) and 160\,$\mu$m (bottom).}
\label{fig:uhya_int}
\end{figure}

The detached shell, which causes the FIR excess emission, has its inner boundary at a distance of 0.11\,pc from the star and a width of 0.02\,pc. Statistical errors on these values are small since the radial intensity profiles fit the respective data from PACS observations well. A comparison for the two PACS bands is shown in Fig.~\ref{fig:uhya_int}. We ought to keep in mind that the shell width is most likely not resolved in the observations and the shape of the profile is additionally broadened by deviations from symmetry. Hence, the derived value must be considered an upper limit. In any case, the dust temperature at the inner shell border is 51\,K.Again, this value has a negligible internal errorbar ($< 1\,$K), but adopting other grain properties (such as changing the provider of the optical constants) will yield different results \citep[cf.][]{Brunner2018}. In Table~\ref{tab:model}, we give an overview of the most relevant model parameters, including their statistical errors, which are internally evaluated by MoD \citep[see][]{Groenewegen2012}.

For the dust mass contained in the detached shell, we derive $(2.2\pm0.4)\times10^{-5}\,\mathrm{M_\odot}$. We assume the MLR does not change within the corresponding time frame, that is, the density follows an $r^{-2}$ distribution. If we adopt the wind velocity of 6.9\,km/s measured in \cite{Olofsson1993a}, the shell width given above translates into a 2800 year period of enhanced mass loss in the case of free expansion (i.e. no wind-wind or wind-ISM interaction). The average dust MLR is then $(7.6\pm1.4)\times10^{-9}\,\mathrm{M_\odot/yr}$. It is, however, not very likely that expansion velocities remain constant during high mass loss events \citep{Mattsson2007}. Examples where kinematic information is available for both present-day mass loss and the expanding detached shell typically show larger velocities of the latter \citep[e.g.][]{Kerschbaum2017}. In general, there is an apparent trend for increasing expansion velocity with higher mass-loss rates, as is, for example, shown in \citet{Bladh2019}. Thus, if we adopt a higher canonical expansion velocity of 15\,km/s, the time frame and MLR change accordingly to 1300 years and $(1.7\pm0.3)\times10^{-8}\,\mathrm{M_\odot/yr}$, respectively. In this free expansion scenario, the rough dynamical age for the shell turns out to be 7500 and 16000 years for wind velocities of 15 and 6.9\,km/s, respectively. The best-fit radial density distribution inside the detached shell is $\propto r^{-1.9}$, which is close to what is expected for a constant MLR. This indicates a quick drop of the MLR after the period of elevated mass loss. The minor deviation might even simply be caused by the arc-shaped emission inside the shell. Based on the density profile and wind velocity from the literature, a present-day MLR of $1.4\times10^{-11}\,\mathrm{M_\odot/yr}$ for the dust and $3.4\times10^{-9}\,\mathrm{M_\odot/yr}$ for the gas (adopting a canonical gas-to-dust ratio of 200) can be estimated as well. The gas/dust ratio might, however, be higher, as stationary wind models show (see Sect.~\ref{discussion}).

\subsubsection{W Ori}

\begin{figure}
\includegraphics[width=0.9\hsize]{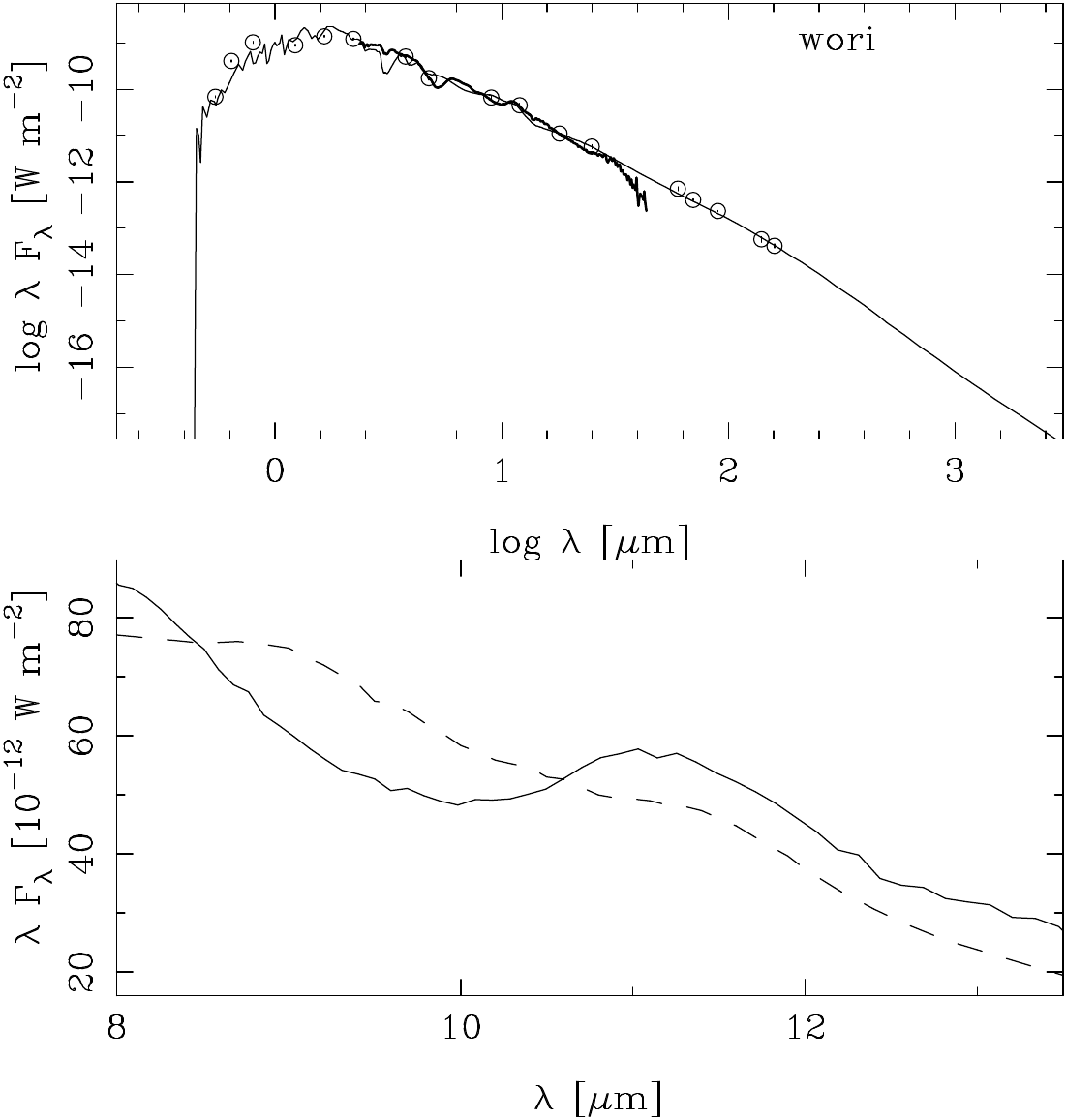}
\caption{Top: SED of the W~Ori MoD model fit compared to the photometric data (for references see Table\,\ref{tab:sed}) and an ISO SWS spectrum. Bottom: Model (dashed) and observed spectrum showing the region of the SiC feature.}
\label{fig:wori_sed}
\end{figure}

\begin{figure}
\includegraphics[width=\hsize]{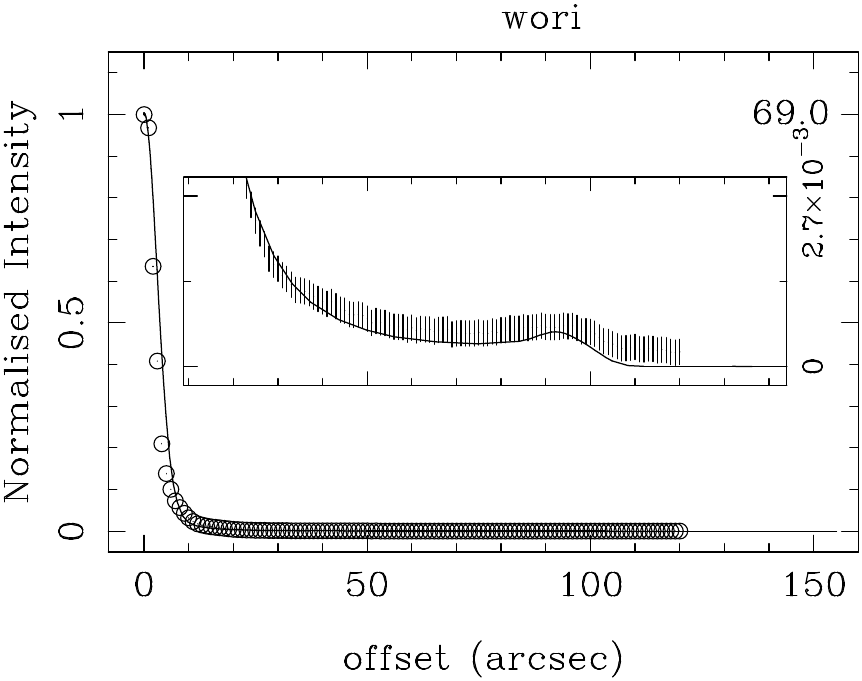}
\caption{MoD model (solid line) and azimuthally averaged radial profile of the PACS observations of W~Ori at 70\,$\mu$m.}
\label{fig:wori_int}
\end{figure}

For the central source, we select a COMARCS model with $T_\mathrm{eff}=2600\,$K and $\mathrm{C/O}=1.1$. The slightly lower C/O (relative to the literature value) was taken because of its availability in the COMARCS grid. The small difference is not expected to critically influence our results, as the changes lie predominantly in spectral features in the near- and mid-IR. There is not much effect on the overall energy distribution, thus, it will not affect our primary goal of describing the cold dust in the envelope. There is an ISO SWS spectrum available for W~Ori that, like in the case of U~Hya, shows the presence of SiC around the star. We thus chose the same dust mixture as for U~Hya. The overall SED, as well as a magnified view of the SiC feature, are given in Fig.~\ref{fig:wori_sed}.
Because of the strong contamination by background sources, we exclude the $160\,\mathrm{\mu m}$ intensity profile from the fitting and only use information from the short wavelength band. The comparison of the best-fit model profile with the 70\,$\mu$m PACS observations is shown in Fig.~\ref{fig:wori_int}. Regarding the error estimate, the considerations in the previous section on U~Hya also apply here.

We arrive at a best-fit model that gives a detached shell dust mass of $(3.5\pm0.3)\times10^{-6}\,\mathrm{M_\odot}$ within a radial range of 0.03\,pc, when adopting a distance of 377\,pc (further model parameters are given in Table~\ref{tab:model}). At the inner boundary, the dust has a temperature of 47\,K, which is very similar to the result obtained for U~Hya. The larger distance from the star and its lower surface temperature compensate the effect of the higher luminosity that is derived for the central source. We again adopt 15\,km/s as the expansion velocity in order to give an estimate on the formation timescales. In the case of a freely expanding wind, the high mass-loss episode would have ended roughly 11000 years ago, having lasted approximately 1600 years. This would correspond to a dust MLR of $(2.2\pm0.3)\times10^{-9}\,\mathrm{M_\odot/yr}$ if we assume a constant mass loss during that period. In comparison, we derive $6.4\times10^{-11}\,\mathrm{M_\odot/yr}$ for the present-day dust MLR when adopting an expansion velocity of 11\,km/s \citep{Schoeier2001}. A drop in that value immediately after the high mass-loss event seems likely given the fact that the slope of the density distribution is close to a value of 2 (see Table~\ref{tab:model}) and shows no major deviations from that point.

\begin{figure}
\includegraphics[width=\hsize]{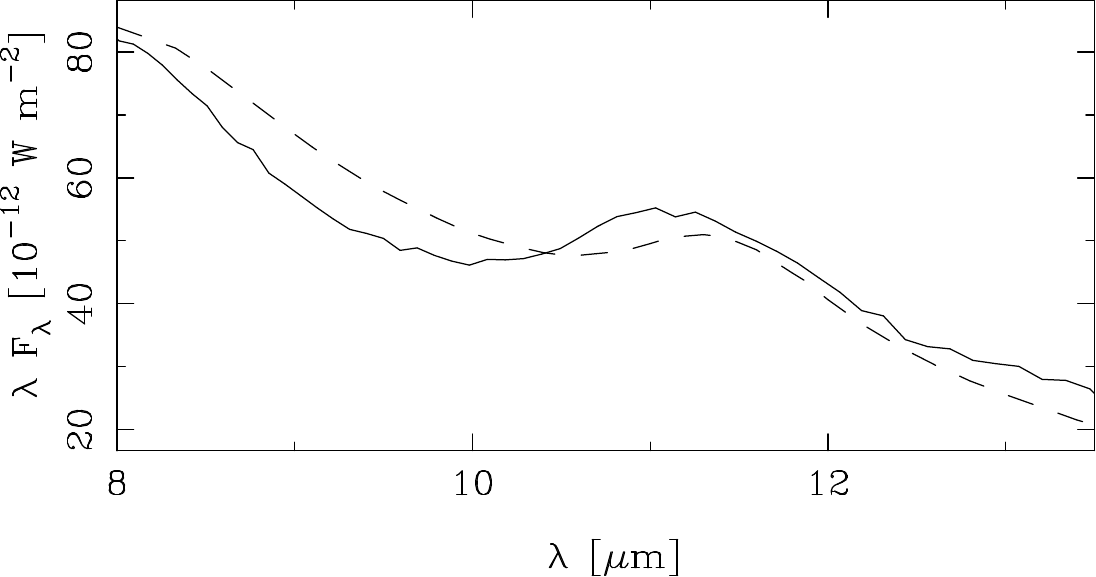}
\caption{SWS (solid line) and MoD model spectrum around the 11.2\,$\mu$m SiC feature. A stellar atmosphere model with $T_\mathrm{eff}=3100$\,K is adopted.}
\label{fig:wori_sicspec}
\end{figure}

While the MoD models reproduce the cold dust emission and the overall SED of W~Ori rather well, a more detailed look at the spectral features reveals pronounced discrepancies. For example, the feature at $11.2\,\mathrm{\mu m}$ is barely present in the model, while it shows up strongly in the SWS data, as can be seen in Fig.~\ref{fig:wori_sed}. Adding more SiC in the present day mass loss would only marginally mitigate the difference. Furthermore, the $\mathrm{C_2H_2}$ absorption at 3\,$\mu$m is much stronger in the model spectrum than it is in the observations. This feature is a good proxy for the effective surface temperature of the star which deepens with lower $T_\mathrm{eff}$ \citep{Paladini2011}. Given the observational evidence, we thus replace the initial input model atmosphere with one with an increased $T_\mathrm{eff}$ of 3100\,K but with otherwise unchanged parameters and re-run the MoD fitting. As it turns out, this modification not only yields an adequate strength of the $\mathrm{C_2H_2}$ feature but also results in a much improved representation of the mid IR spectrum around the SiC emission. The latter can be nicely recognised when comparing the spectra in Figs.~\ref{fig:wori_sed} and \ref{fig:wori_sicspec}. With the adapted model, the derived dust quantities are somewhat altered:\ the present-day dust MLR increases to $7.0\times10^{-11}\,\mathrm{M_\odot/yr}$ and the dust mass in the detached shell to $3.9\times10^{-6}\,\mathrm{M_\odot}$. The dust temperature at the inner boundary of the detached shell shifts to 48\,K (cf. Table~\ref{tab:model}).

\begin{table*}[h]
\centering
\caption{Best-fit model parameters obtained from \textit{MoD} calculations, including statistical errors.}
\label{tab:model}
\begin{tabular}{lrrcrrrrrc}
\hline
\hline
\vspace{-9pt}\\
 & $L\,[\mathrm{L_{\sun}}]$ & $\tau_{0.55\mu\mathrm{m}}$ & $T_\mathrm{d}\,[\mathrm{K}]$ & $p_1$ & $y_1\,[\mathrm{pc}]$ & $\delta y\,[\mathrm{pc}]$ & $s_1$ & $p_2$ & $M_\mathrm{d}\,[\mathrm{M_{\sun}}]$\\
\hline
\vspace{-8pt}\\
U~Hya & $5670\pm50$ & $0.09\pm0.001$ & $51$ & $1.9\pm0.007$ & $0.11\pm0.001$ & $0.02\pm0.001$ & $230\pm12$ & $2.0$ & $(2.2\pm0.4)\times10^{-5}$\\
W~Ori & $9500\pm180$ & $0.28\pm0.003$ & $47$ & $2.05\pm0.003$ & $0.17\pm0.007$ & $0.03\pm0.004$ & $15\pm1$ & $2.0$ & $(3.5\pm0.3)\times10^{-6}$\\
\hline
\end{tabular} 
\tablefoot{$L$ is the luminosity of the central star, $\tau$ is the optical depth of the entire dust envelope at $0.55\,\mu\mathrm{m}$ and $y_1$ and $\delta y$ are the radius and width of the detached shell, given in pc. The condensation temperature $T_\mathrm{c}$ was kept fixed at 1000\,K. $T_\mathrm{d}$ is the dust temperature at distance $y_1$ and $M_\mathrm{d}$ is the dust mass in the detached shell with thickness $\delta y$. $p_1$ and $p_2$ are the slope of the density power law for the inner part of the shell and the detached shell, respectively. $s_1$ is the density contrast of the detached shell. The provided formal errors are the variation of the parameters which would yield a $\chi^2=1$, including all observational constraints.}
\end{table*}

\subsection{Stationary wind models}

The stellar component in the wind model is again a hydrostatic atmosphere, where we now adopt a newly calculated COMARCS model with parameters tailored to U~Hya. Its inherent self-consistent luminosity is sufficiently close to what is obtained in the MoD results: $5500\,\mathrm{L_\odot}$, compared to the previously fitted $5670\,\mathrm{L_\odot}$. Also the other stellar parameters, except for an elevated C/O, are representative of literature values of U~Hya. 

The grid variables in Table~\ref{tab:coma_grid} determine two physical quantities of interest, namely the present-day dust MLR and the dust mass contained in the detached shell. 
The most fundamental of these parameters is the pdMLR, which determines (in combination with the initial outflow speed $u_\mathrm{ini})$ the present-day dust MLRs for amC and SiC that are also listed in the table. The detached shell dust mass, on the other hand, depends on the density scale, $\Delta \rho$, the shell width, $\Delta r_\rho$, its position, $r,$ and also the dust pdMLR. In the following, we present the way the modelled observables, that is, the SED and intensity profiles, behave when those parameters are changed.

\begin{figure}
$
\begin{array}{c}
\includegraphics[width=\hsize]{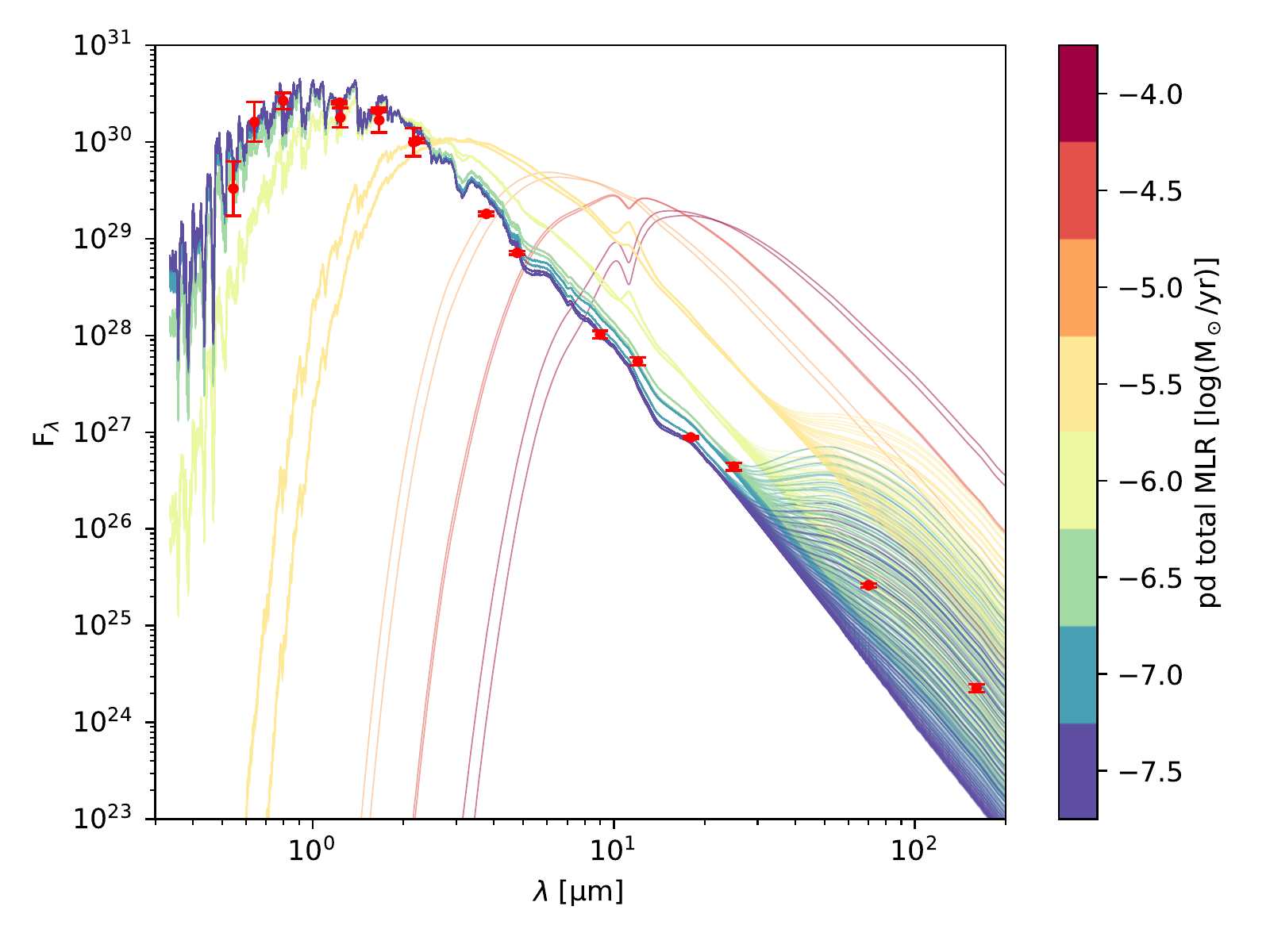} \\
\includegraphics[width=\hsize]{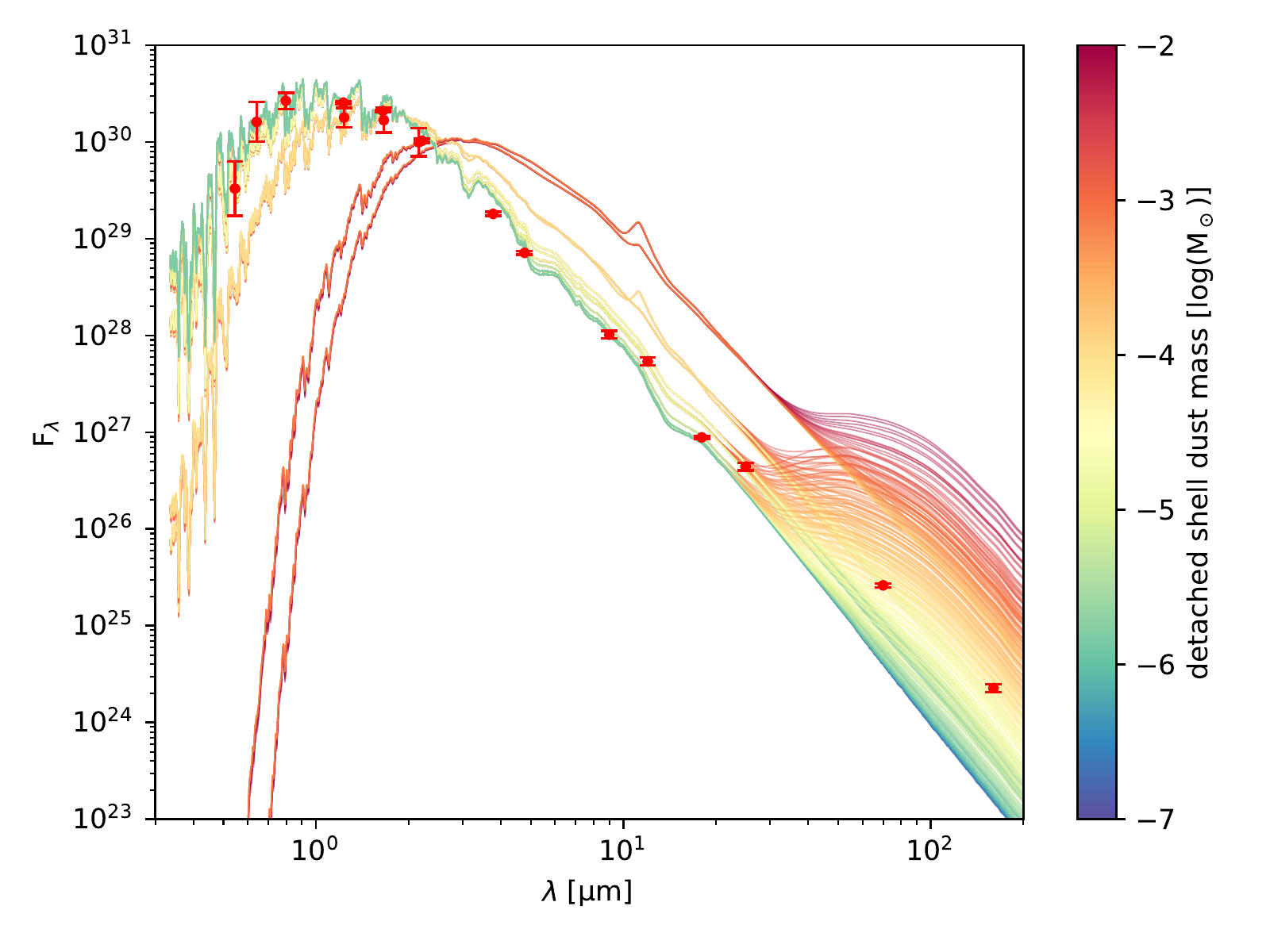} 

\end{array}$
\caption{Model spectra based on combined stationary wind models for U~Hya. The line colours indicate the total present day mass-loss rate (top) and the variation in total mass contained in the detached shell (bottom). Red symbols are photometric observations with their respective error bars.}
\label{fig:uhya_spec_gail}
\end{figure}

\begin{figure}
$
\begin{array}{c}
\includegraphics[width=\hsize]{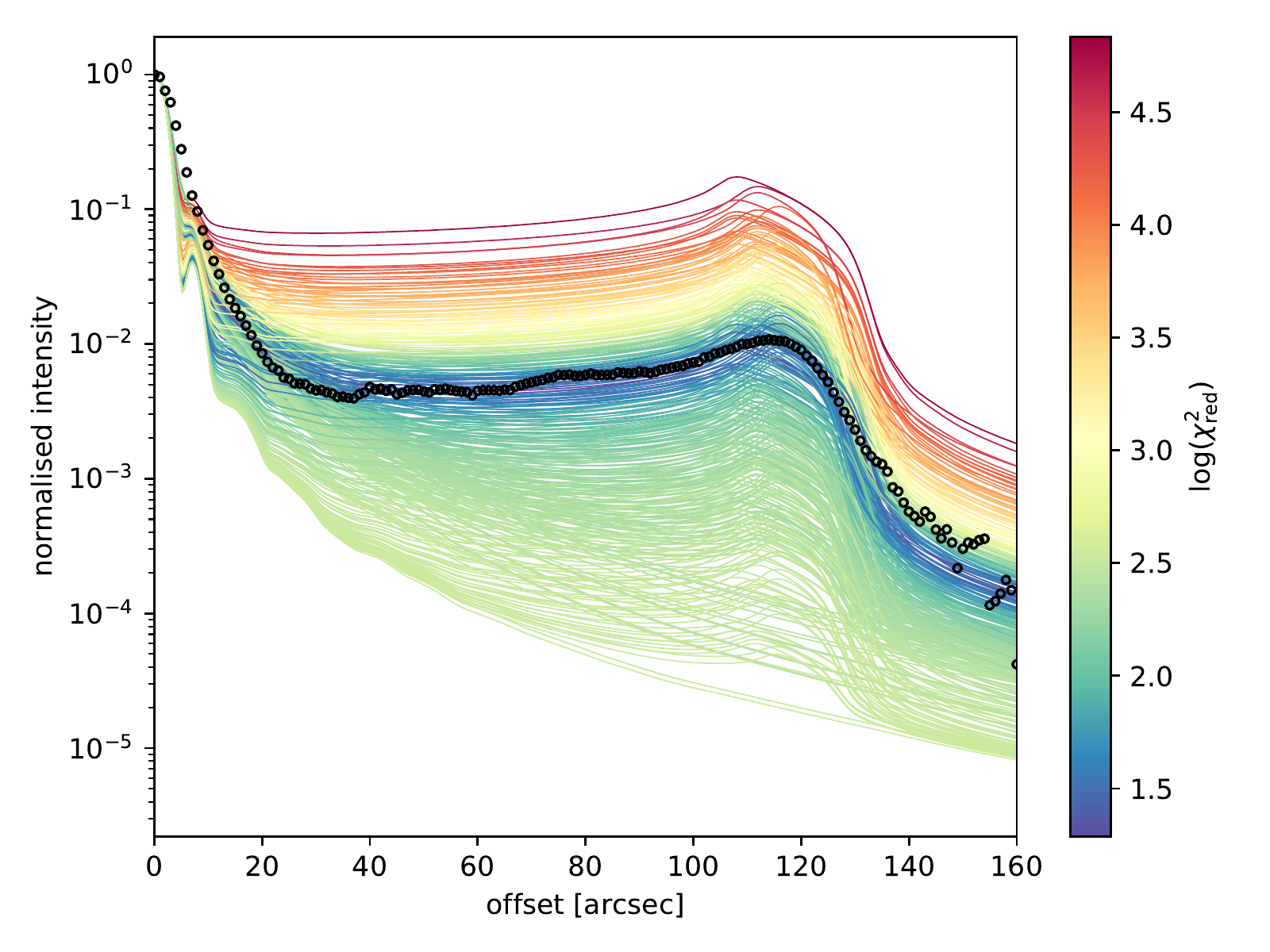} \\
\includegraphics[width=\hsize]{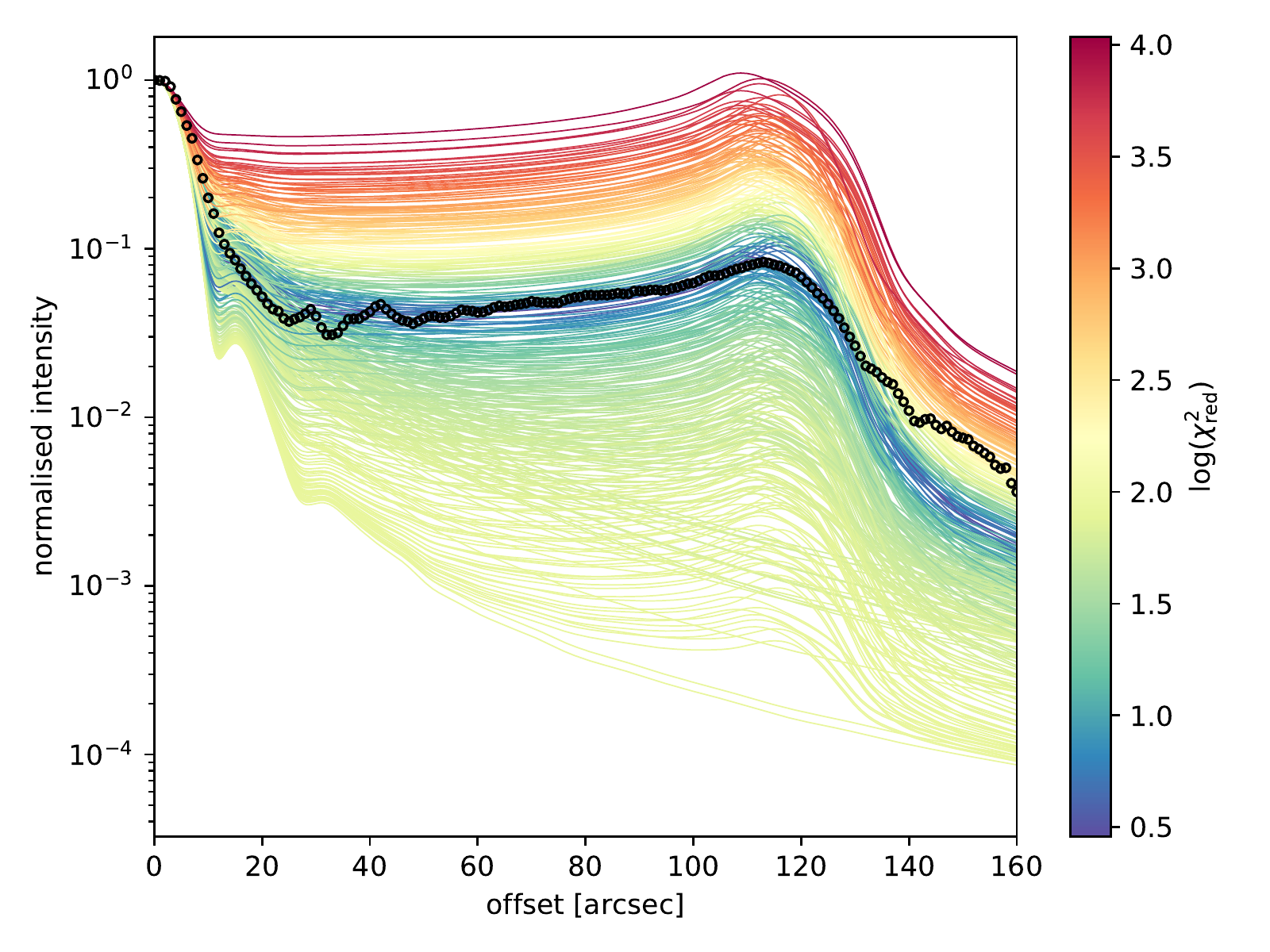}
\end{array}$
\caption{Radial brightness profiles calculated from the stationary wind models at 70 (top) and $160\,\mathrm{\mu m}$ (bottom). The colours reflect the goodness of the fit to the respective observed PACS brightness distribution (black open circles).}
\label{fig:uhya_int_gail}
\end{figure}

The present-day mass loss represents warm dust at a distance of several stellar radii. Thus a variation in the pdMLR is expected to primarily show an impact in the optical and near-to-mid-IR region of the SED. As can be seen in Fig.~\ref{fig:uhya_spec_gail}, this is indeed the case when a growing MLR increasingly suppresses the flux at shorter wavelengths. While the intensity of absorption by dust particles is moderate for intermediate mass-loss rates, the flux at short wavelengths plummets with more extreme values. In the latter cases (typically for total MLRs $\gtrsim 10^{-5}\,\mathrm{M_\odot/yr}$), a star would be basically undetectable in optical bands. Such heavy mass loss is found to occur only at the very end of AGB evolution, also known as the superwind phase \citep[see e.g.][]{Lagadec2008}. Observations clearly suggest a rather low pdMLR for U~Hya since the models with high MLRs  show in Fig.~\ref{fig:uhya_spec_gail} large deviations from the overall energy distribution defined by the photometric measurements.

The dust in the detached shell is located at several thousand stellar radii and is, thus, much cooler compared to the present day ML. According to Wien's displacement law, a blackbody with a temperature of 50\,K (the dust temperature derived for U~Hya in the previous section) has its emission peak at $\sim 60\,\mathrm{\mu m}$. As expected, this is exactly where we see excess emission in our model spectra (bottom graph in Fig.~\ref{fig:uhya_spec_gail}). The intensity of this bump correlates well with the amount of dust we put in the detached shell. Again, it is apparent that the chosen parameter grid covers a wide range of possible configurations. It is obvious that only some of them agree with the photometric observations. Adding a detached shell to models with an already extreme present-day total MLR ($\geq 10^{-5}\,\mathrm{M_\odot/yr}$) results in an unreasonably high total amount of material in the detached shell. Thus, for the concerned MLRs, we only include the baseline models for illustrative purposes.

The radial flux distribution derived from the two Herschel/PACS maps allows us to further spatially constrain the origin of the FIR emission. In Fig.~\ref{fig:uhya_int_gail}, we compare the observations with intensity profiles derived from our wind models. We convolve the raw model profiles with an azimuthal average of a synthetic point spread function of the respective PACS cameras. Nevertheless, owing to the complex nature of the PSF, one can only approximate the real situation. This primarily concerns  the innermost regions ($\lesssim20\arcsec$), that is, the present-day mass loss. Moreover, outside the detached shell the models underestimate the observed intensity. First, this comes from the broadening of the observed profile caused by the slight asymmetries of the detached shell. Secondly, the models do not include the background emission that increases the basic flux level, which becomes particularly apparent in the outer regions of the long wavelength channel. Overall, however, the observations can be well reproduced by a narrow range of grid models in both the 70 and $160\,\mathrm{\mu m}$ bands.

We separately evaluate the goodness-of-fit of the individual models with respect to photometric filters and the intensity profiles in terms of $\chi^2_\mathrm{red}$, where
\[ \chi^2_\mathrm{red}=\dfrac{\chi^2}{n-1-f}, \]
$\chi^2$ is defined in Eq.~\ref{eq:chi2}, $n$ is the number of sample data points, and $f$ is the number of free parameters. The photometric data are from the sample used with MoD (see Table~\ref{tab:sed}) including the respective errorbars. For the data points of the PACS intensity profiles, we adopt a relative error of 5 and 10\% for the short and long wavelength band, respectively. Going forward, we only consider data points up to a{ 140\arcsec} radius, which is sufficient to include the whole detached shell. This discards the outer regions that are not properly accounted for by the models.

\begin{figure}
\includegraphics[width=\hsize]{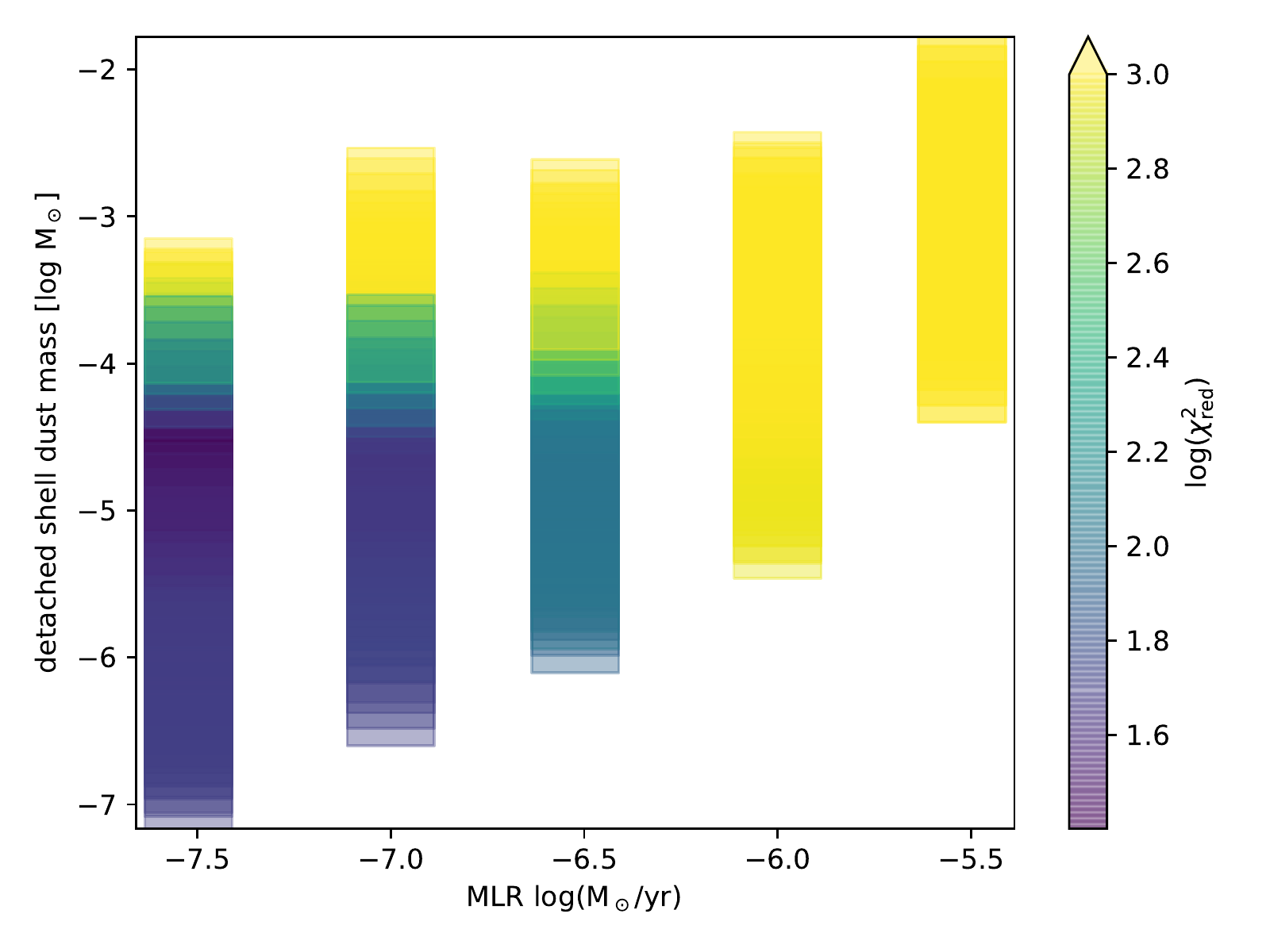}
\caption{$\chi^2_\mathrm{red}$ values of the stationary wind model grid, evaluated based on photometry.}
\label{fig:chi2map_phot}
\end{figure}

\begin{figure}
$
\begin{array}{c}
\includegraphics[width=\hsize]{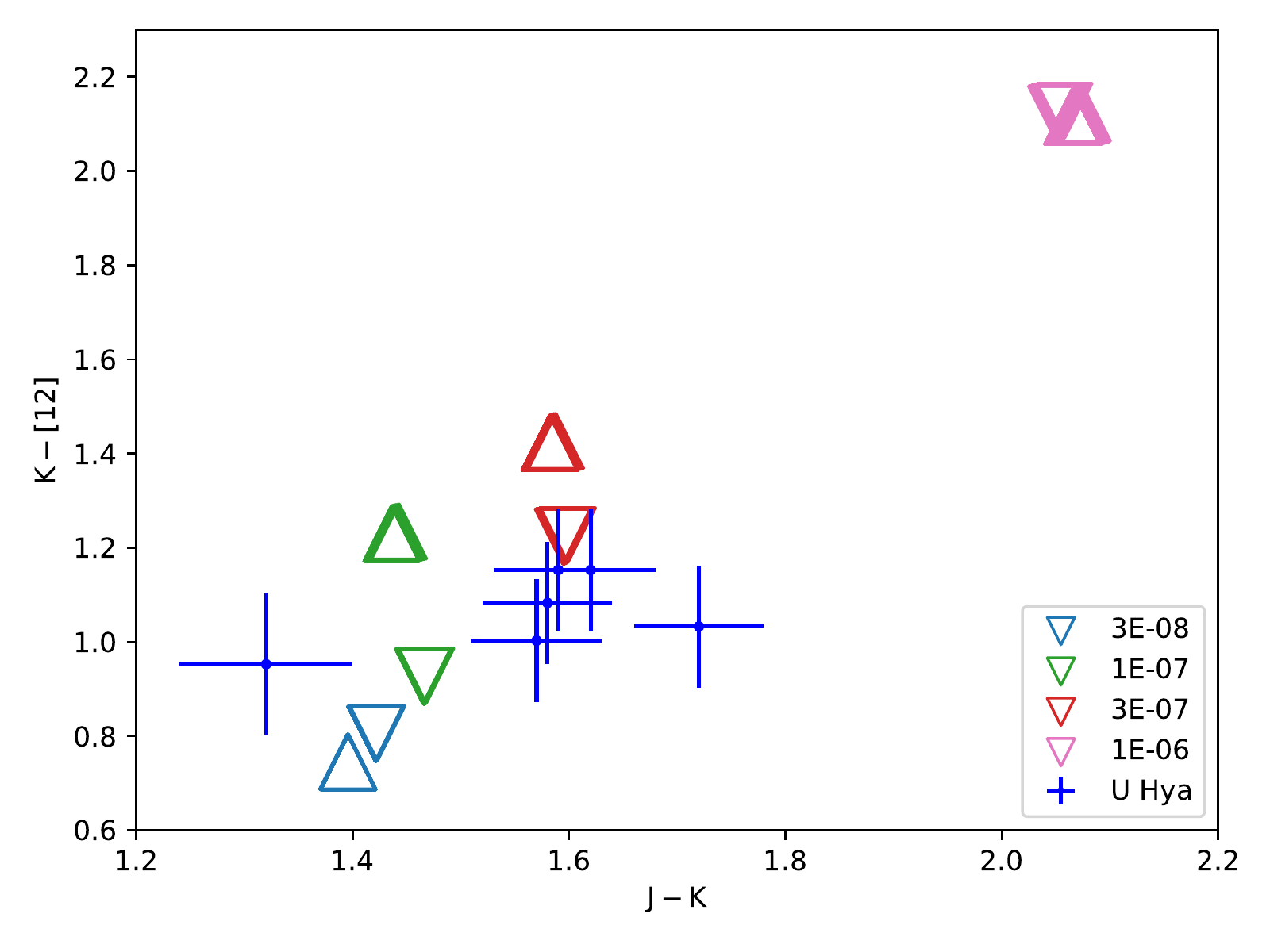} \\
\includegraphics[width=\hsize]{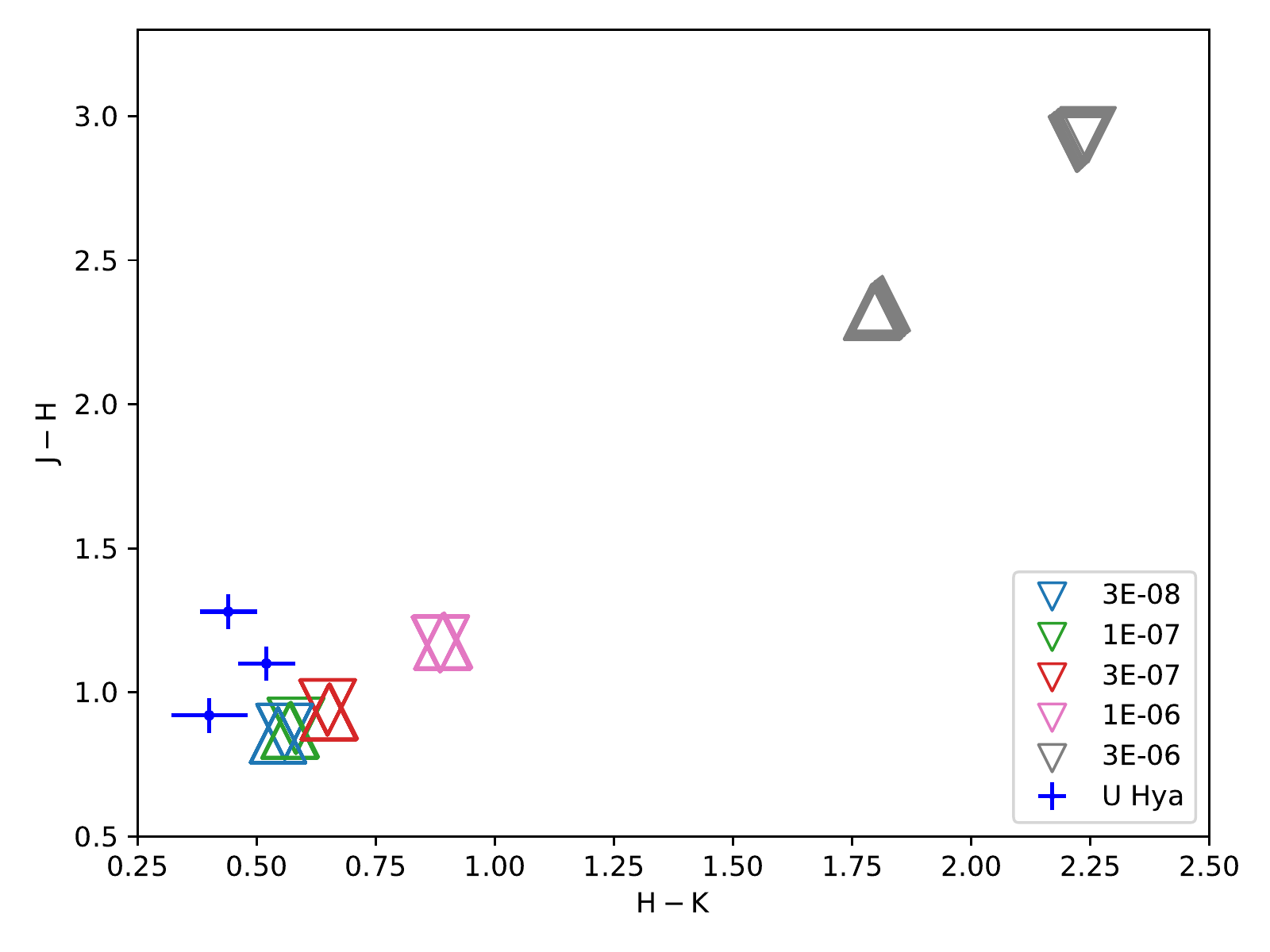}
\end{array}$
\caption{Colour-colour diagrams of U~Hya derived from the stationary wind models. The coloured triangles represent the individual models with their respective present day total mass-loss rates (in $\mathrm{M_\odot/yr}$). The upright triangles are models with high, the downward facing ones with low initial outflow velocity (see text). Blue dots with errorbars indicate observed values from the literature (see Table~\ref{tab:sed}).}
\label{fig:uhya_ccd_gail}
\end{figure}

When comparing the synthetic photometric fluxes to the observational data, it becomes clear that models with a low present-day MLR are favoured. Figure~\ref{fig:chi2map_phot} shows a $\chi^2$ analysis for the model grid. As is already evident from the model spectra in Fig.~\ref{fig:uhya_spec_gail}, winds with a total pdMLR in excess of $10^{-6}\,\mathrm{M_\odot/yr}$ obviously fail to reproduce the optical and NIR fluxes regardless of the other grid properties. The best fits are obtained for a pdMLR of $3\times10^{-8}\,\mathrm{M_\odot/yr}$, where the detached shell density is typically enhanced by a factor of 1000, depending on the corresponding shell width. Also, models with a total pdMLR of $10^{-7}\,\mathrm{M_\odot/yr}$ but a lower contrast in the detached shell of a few 100 fit the photometry well. In both cases, the dust mass contained in that shell is around $3\times 10^{-5}\,\mathrm{M_\odot}$. Generally, while the dust mass in the detached shell is relatively consistent for the best fit models, there is a degeneracy in the parameters that constitute this quantity, meaning that shell width (which is, most likely, not resolved in the PACS maps) and density scale cannot be well-constrained. This is not surprising given that for a certain dust temperature that is well-confined by the shell peak position in the radial profiles, it is only the amount of dust that matters in an optically thin regime. 
The fact that the best results are obtained with models having the lowest pdMLR available in the grid raises the question of whether even lower values -- or no present-day mass loss at all -- would further improve the photometric fit. On the other hand, the [$K-12$] versus [$J-K$] colour-colour diagram, presented in Fig.~\ref{fig:uhya_ccd_gail}, suggests the presence of a non-negligible amount of warm dust, corresponding to a total pdMLR in the range of $(1-3)\times10^{-7}\,\mathrm{M_\odot/yr}$; whereas in the $[J-H]$ versus $[H-K]$ diagram, none of the models can adequately represent the observations. This issue is discussed in the next section.

\begin{figure}
\includegraphics[width=\hsize]{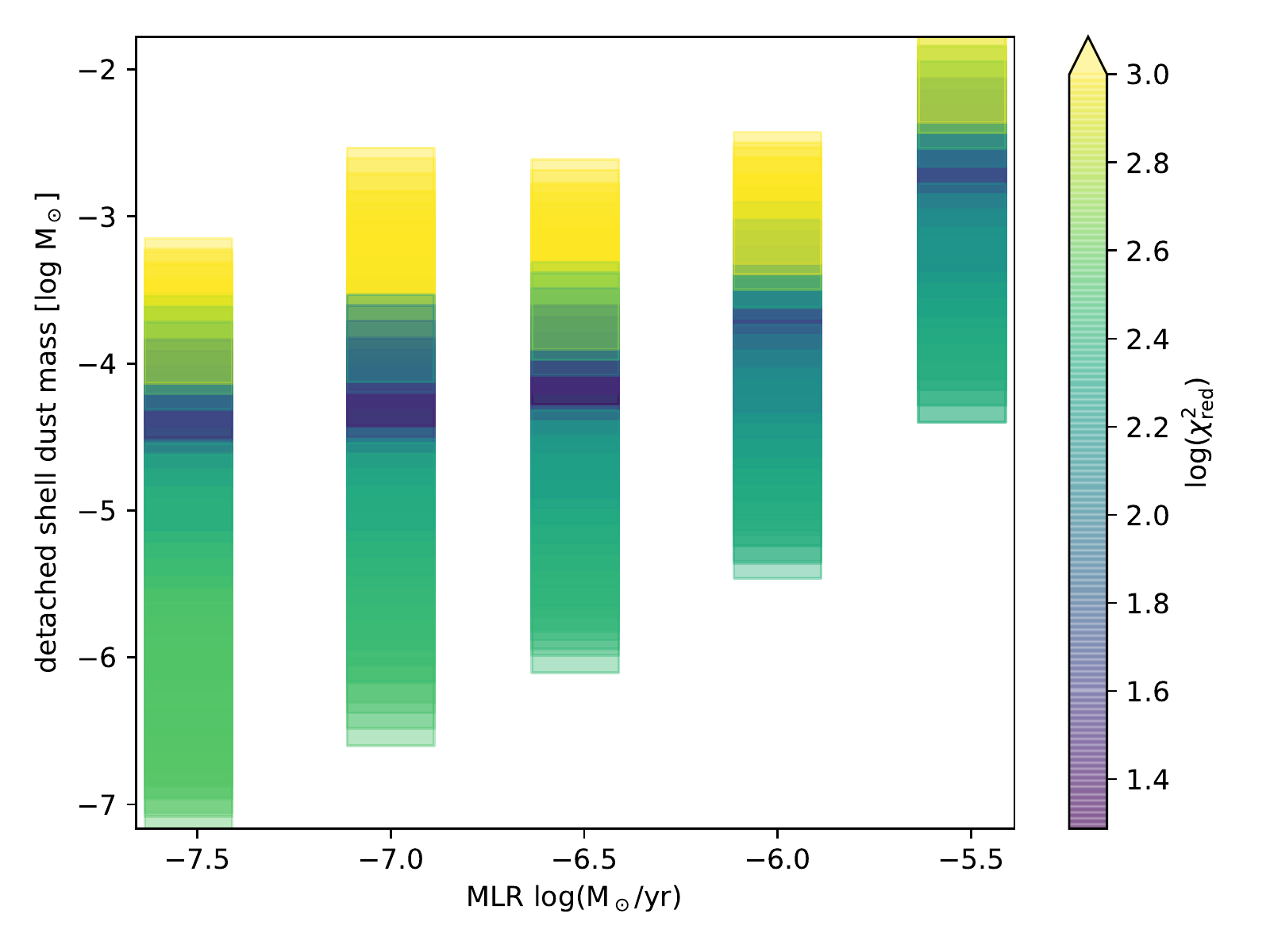}
\caption{$\chi^2_\mathrm{red}$ values of the stationary wind model grid, based on the 70\,$\mathrm{\mu m}$ radial profiles. Data for the 160\,$\mathrm{\mu m}$ profile are very similar.}
\label{fig:chi2map_prof}
\end{figure}

Compared to the photometric evaluation, the intensity distributions (Fig.~\ref{fig:chi2map_prof}) are not as sensitive to the pdMLR and do not strictly exclude higher values. The $70\,\mathrm{\mu m}$ band tends to favour a total pdMLR of $10^{-7}\,\mathrm{M_\odot/yr}$ or $3\times 10^{-7}\,\mathrm{M_\odot/yr}$. On the other hand, the dust mass in the detached shell is better constrained, suggesting $8\times 10^{-5}\,\mathrm{M_\odot}$, typically with a density enhancement by a factor of 250. Here, the two PACS profiles are in good agreement.

\section{Discussion}
The spatially well resolved image data prove to be a valuable additional constraint regarding the dust density distribution for the 1D models. Our mass estimates obtained from the two modelling approaches yield consistent figures and are generally in line with other studies. Some aspects of the spectral results, however, reveal some discrepancies between our models, observations, and literature.

\subsection{U Hya}

\label{discussion}

\subsubsection{Geometry}
Based on our PACS maps, we measure the location of the shell peak emission at a radial distance of 114\arcsec or 0.12\,pc. Under the assumption that the extensive instrument point spread functions most likely determine the observed width of the structure, the approximate position we find for the inner boundary of the detached shell is in agreement with the results from \citet{Izumiura2011}. These authors derive a range of $101\arcsec \mbox{--} 107\arcsec$ by fitting radial intensity profiles to AKARI imaging data. For the shell width, they give $16\arcsec \mbox{--} 23\arcsec$. With the better-resolved Herschel data we are not able to significantly further narrow down these values when analysing the PACS maps. 

The results from the MoD models are compatible with a detached shell width within this range, but a narrower shell can reproduce the observed data as well. Similarly, the grid of stationary wind models shows no clear preference regarding the shell width parameter and even the lowest grid values ($\sim 13\arcsec$ or 0.013\,pc) yield results that agree with the observations. Thus, an even (geometrically) thinner structure cannot be excluded by either the MoD nor the wind models and the best-fit values have to be considered an upper limit at best. Partial responsibility for this rather high limit (compared to the PSF FWHM) comes from the slight asymmetries and inhomogeneities (Fig.~\ref{fig:uhya_polar}) of the shell as these properties broaden the shell feature in the radial profiles and, therefore, hamper efforts to obtain a better constraint on the width. The realistic possibility of a much more spatially confined detached shell is supported by highly resolved sub-mm interferometric maps of similar objects. In the case of R~Scl \citep{Maercker2016}, and particularly U~Ant \citep{Kerschbaum2017}, the structures detected in CO line emission appear to be only a few arcseconds thin. However, the dust does not necessarily need to be exactly aligned with the gas component and could show a more diluted distribution.

The probable formation mechanism behind the detached shell is the interaction of a faster, denser wind -- which develops as a consequence of a thermal pulse -- with the preceding outflow. This is supported by the detection of $^{99}$Tc in the star, indicating a recent thermal pulse event, in whose aftermath this relatively short-lived element is brought to the stellar surface. 

\begin{figure}
\includegraphics[width=\hsize]{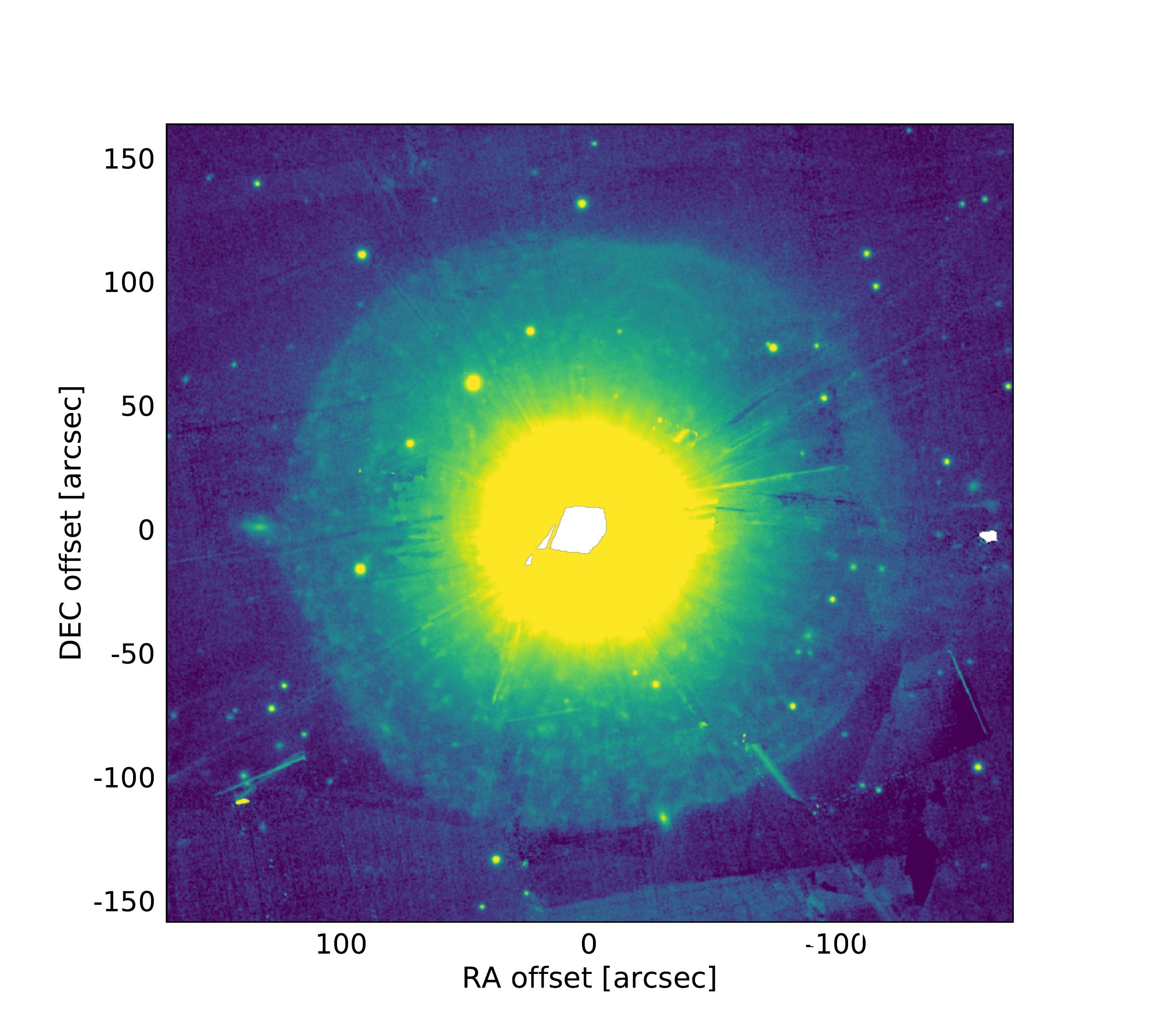}
\caption{PANSTARRS scattered light image (\textit{r} filter) of U~Hya.}
\label{fig:uhya_panstarrs}
\end{figure}

The alternative ISM interaction scenario is less likely, as already discussed by \cite{Izumiura2011}, mainly owing to the low interstellar densities and the overly compact size of the shell. In its current evolutionary state, ISM influence is not expected to be chiefly responsible for the observed density enhancements around U~Hya \citep{Villaver2002}. There is, however, morphological evidence of some interaction of the stellar wind with the surrounding medium. For one, the star is displaced from the shell center in good alignment with the direction of its space motion (see Sect.~\ref{uhya_morph}). Moreover, at a very similar position angle, the detached shell is slightly oblate. This deviation from spherical symmetry as well as the overall detached shell structure is also traced in optical scattered light observations by \cite{Izumiura2007} and in the image catalogue of the \textit{PANSTARRS} project \citep[Fig.~\ref{fig:uhya_panstarrs},][]{Chambers2016}. At even shorter wavelengths, \cite{Sanchez2015} recently found far-UV emission in GALEX data, co-located with the thermal dust emission. The FUV radiation preferentially originates from the region where the ISM headwind is expected to most intensely interact with the detached shell and, thus, they argue that shocks at the wind/ISM interface are the most probable cause.

To our knowledge, U~Hya remains a singular case in the way it shows such consistent evidence of the potential influence of space motion on the geometry of detached shells, at least within the small known sample of this class of targets. Whereas, for TT~Cyg, \citet{Olofsson2000} also find a displacement between the shell centre and the star, the offset is not aligned with the stellar space motion and, hence, is more likely caused by a binary companion. Moreover, for some objects such as S~Sct, there are indications in FIR data \citep{Mecina2014} for some minuscule effects of space motion, but only on the very extended and diffuse structure outside the main shell. Besides the group of detached shell targets, there are, of course, several AGB stars in the MESS sample alone that show blatant signs of wind-ISM interaction as they plough through the surrounding matter at high speeds \citep[see, e.g. the `fermata' class in][]{Cox2012}. Furthermore, \citet{Randall2020} recently observed a spiral structure around the O-rich AGB star \object{GX~Mon}, where there seem to be indications of potential deformation due to the space motion of the source.

\subsubsection{Mass loss}
\label{discussion_uhya_massloss}
Results based on our stationary wind models suggest a current gas MLR between $3\times10^{-8}$ and $3\times10^{-7}\,\mathrm{M_\odot/yr}$, depending on whether we are considering the SED or the intensity profile. As the SED is more sensitive to the pdMLR (Figs.~\ref{fig:chi2map_phot}\,and\,\ref{fig:chi2map_prof}), the lower values that are suggested by the photometric fit are more likely.

We also create synthetic colour-colour diagrams for U~Hya from the stationary wind results (Fig.~\ref{fig:uhya_ccd_gail}) as an additional constraint for the models. For example $[K-[12]]$ vs $[J-K]$, which is particularly sensitive for the warm dust and thus the present day mass loss. Here, the observational data is in agreement with the models with a total pdMLR of $(1-3)\times10^{-7}\,\mathrm{M_\odot/yr}$. Within that set of models, the ones with lower initial gas outflow speeds are favoured. Moreover, it can clearly be seen that the detached shell mass has effectively no influence at near to mid IR wavelenghts, as the respective model sets virtually show no scatter. Our MLR range is in agreement with \citet{Olofsson1993a}, who observed unresolved molecular line emission around the star and derive a corresponding present-day mass-loss rate of $1.2\times10^{-7}\,\mathrm{M_\odot/yr}$ with an outflow velocity of 6.9\,km/s.

For the present-day gas MLRs, our wind models show a gas-to-dust ratio on the order of $10^3$ after dust growth has effectively stopped, which typically occurs around $10^{14}$\,cm for the low MLR models (see Fig.~\ref{fig:density_profs}). Corresponding to the best-fit range given above this translates to dust pdMLRs between $1.8\times10^{-12}$ and $4.3\times10^{-10}\,\mathrm{M_\odot/yr}$ (see Table~\ref{tab:coma_grid}). With MoD, we derive similar values: $1.4\times10^{-11}\,\mathrm{M_\odot/yr}$ adopting DHS grain geometry, and $3\times10^{-11}\,\mathrm{M_\odot/yr}$ when instead using solid spheres of the same size. For smaller grain radii, the MLR would further increase to $\sim~5\times~10^{-11}\,\mathrm{M_\odot/yr}$. With the DHS properties used in this paper, the typical mass difference to solid spheres is, hence, about a factor of between 2-3, as was also found by \cite{Brunner2018}, who used comparable grain geometries in their study. Similarly, we obtain a smaller dust mass of the detached shell of $2.2\times10^{-5}\,\mathrm{M_\odot}$ from the MoD best fit (using DHS), compared to the values derived from the stationary wind models. In that case, $3\times10^{-5}\,\mathrm{M_\odot}$ and $8\times10^{-5}\,\mathrm{M_\odot}$ are obtained when evaluating for the SED or radial intensity profiles, respectively. Considering that the MoD figure takes into account both SED and radial profiles, a compromise between the latter two wind model values would be in line with what is to be expected from the different grain geometry. All of these numbers are, however, still a bit below the $(0.9-1.4)\times10^{-4}\,\mathrm{M_\odot}$ that was found by \cite{Izumiura2011} from intensity profile modelling based on AKARI data for a distance of 162\,pc. 

Concerning the observed colours one has to keep in mind that they are obtained from photometry which was taken at different epochs and thus there is a spread in the data points (mostly due to variations in the J-band), extending the parameter range of consistent models. Nevertheless, such a plot demonstrates how strong the reddening increases beyond $3\times10^{-7}\ \mathrm{M_\odot/yr}$ and clearly excludes values higher than that. This is in line with the findings from the photometric fit (Figs.~\ref{fig:uhya_spec_gail}\,and\,\ref{fig:chi2map_phot}). By comparison, in $[J-H]$ vs $[H-K]$ models are not able to reproduce the observed colours equally well. While for the $[J-H]$ colour models with gas pdMLRs up to $10^{-6}$\,M$_\odot$/yr lie in the observed range, there appears to be a systematic offset between models and observations in the $[H-K]$ colour.   This is mainly due to incomplete C$_2$ opacity in the COMARCS models. Better opacity data from the \emph{ExoMol} project \citep{Yurchenko2018} will improve the situation and move the models into the region of the measured data.

Finally, we would like to comment on the adopted C/O ratio. The high C/O of 2 (instead of 1.04, as suggested by observations in Table~\ref{tab:targets}) is used in the stationary wind models for practical reasons. It is a necessary condition in order to get a substantial matter outflow. Moreover, dynamical models also show that intense dust-driven winds only arise for higher C/O values \citep{Eriksson2014,Bladh2019}. In addition, with a C/O of 1.04, a very high gas MLR would be needed ($\sim25$-fold) to have enough free carbon available to form the observed amounts of amC dust. Also from an observational perspective, there are indications that adopting a high C/O ratio is justified, as we briefly discuss at the end of section \ref{discussion_wori}. A high value of 2.38 is found by \citet{Rau2017} for U~Hya as well. The impact of the assumed C/O on the result of the MoD fits remains small since in the range around 3000\,K, it has no strong effect on the overall energy distribution of the central sources.

\subsection{W Ori}
\label{discussion_wori}
Contrary to U~Hya, there was no previous observational evidence for extended circumstellar structures around W~Ori. Based on unresolved CO line observations, \cite{Schoeier2001} derive a present day mass-loss rate of $7 \times 10^{-8}\,\mathrm{M_\odot/yr}$ (for a distance of 220\,pc) and an expansion velocity of 11.0 km/s. Although the atmospheric C/O ratio of 1.16 found by \cite{Lambert1986} might point to a rather evolved state of the object and thus to potentially considerable amounts of matter already being expelled by the stellar wind, only Herschel/PACS observations reveal a very thin, spherically symmetric detached shell that points to a past episode of elevated mass loss. We don't detect any apparent signs of interaction of the stellar wind with the surrounding ISM. Given the low space velocity (Table~\ref{tab:kinematics}), also compared to U~Hya, this is not surprising. We derive a dust mass of $(3.5\pm0.3)\times10^{-6}\,\mathrm{M_\odot}$ contained in the detached shell, which is about an order of magnitude lower than what we obtain for U~Hya. Also the density contrast between the detached shell and the present day mass loss is lower by a similar factor. However, in view of the potentially underestimated distance (see Gaia measurements) we stress that the derived mass might be too low. Likewise, the shell's spatial extent and estimated formation timescale would be affected. In any case, the low densities are a possible explanation for the lack of detected molecular line emission. At such large distances from the star, CO molecules are expected to be effectively dissociated by energetic photons from the interstellar radiation field \citep[see e.g.][]{Saberi2019}. We could, in principle, look for the products of this reaction, for example, CI, to trace the gas component, as was shown by \cite{Olofsson2015} for the case of R~Scl. However, for the targets presented in this paper, given their angular size and weak flux, this is impractical regarding the required observation time at facilities, such as ALMA, that would provide the necessary spatial resolution.

The ISO SWS spectrum shows a deep absorption feature around $5\,\mathrm{\mu m}$ that is ascribed to C$_3$ and is also not fully reflected in the COMARCS model atmospheres. It has been suggested that the feature depth must correlate with a particularly high C/O ratio \citep[\mbox{$\sim2$},][]{Gautschy-Loidl2004,Aringer2019}. This is much higher than what is found by \cite{Lambert1986} and used for the input radiation source in our model. More interestingly, all known detached shell objects -- all of them carbon stars -- for which spectral data of that region are available share this pronounced dip in the spectrum and are thus assumed to share a similarly high C/O ratio. In this view, also the C/O value of 1.04 obtained for U~Hya, for which no SWS data are available, seems rather low. As we mention in Sect.~\ref{discussion_uhya_massloss} it will be very difficult to explain the observed amount of dust and almost impossible to obtain a dust-driven wind if C/O remains too close to one (< 1.1--1.2). Nevertheless, \citet{Abia2015} also list very low values for U~Hya (1.05) and W~Ori (1.07). This difference between results based on some high-resolution measurements and the requirement to get a strong C$_3$ feature and a considerable amount of dust requires some further investigation. Such a study must include a consistent treatment of complete molecular opacities in the model construction and spectrum synthesis as well as possible dynamical and non-equilibrium effects \citep[see][]{Aringer2019}.

\section{Conclusion}
We present FIR Herschel/PACS observations of the two carbon stars U~Hya and W~Ori. Around both targets, we detect thermal dust emission originating from geometrically thin, spherically symmetric shells. While the data for U~Hya confirm the results from previous IR space missions and refine the morphological picture, the extended structure surrounding W~Ori was first detected within the MESS programme, further increasing the number of objects known to host a detached shell. We argue that those shells formed during a short period of enhanced mass loss following a recent thermal pulse, which was potentially supported by wind-wind interaction with a slower outflow.

For both sources, we calculate radiative transfer models of the circumstellar dust envelope adopting a parametrised density distribution. The derived values for the dust mass, temperature, and spatial scales are in line with what is commonly found for other examples of the detached shell sample. The spatial constraints provided by the PACS maps critically help to determine the shell density profiles and, thus, the recent mass-loss history of the respective objects. In the present cases, our models suggest a rather abrupt decrease in MLR after the high mass-loss phase, as has been predicted by stellar evolution models. Since the detached shells around U~Hya and W~Ori have, so far, not been detected in line emission, no kinematic information is available and we can only speculate about the involved timescales. Yet, given the quite narrow range of expansion velocities found for AGB outflows, adopting canonical values at least allows for reasonable estimates.

In addition to the parametrised radiative transfer, we are also able to reproduce the observational data of U~Hya by means of stationary wind models. They provide a more complete description of the circumstellar structure and allow us, for example, to get an estimate of the gas component in the shell as well. Concerning the derived dust budget, the two model approaches yield similar results.

In view of the results presented in this paper, it is clear that the spatially resolved data of thermal dust emission are a critical constraint when modelling the extended matter distribution around stars. Moreover, when looking at the presently known sample of detached shell objects, for the spatially more extended (i.e. dynamically, most likely older structures) FIR to sub-mm continuum observations are the preferred method of detection in considering the photodissociation of CO at large radii. 

\begin{acknowledgements}
MM and FK acknowledge funding by the Austrian Science Fund FWF under project number P23586 and FFG grant FA538019. BA was supported by the ERC Consolidator Grant funding scheme ({\em project STARKEY}, G.A. n.~615604). MG acknowledges support by an ESA-Prodex grant.
\end{acknowledgements}

\bibliographystyle{aa}
\bibliography{pap3}

\end{document}